\documentclass[a4paper,12pt]{article}
\usepackage{amsmath,amssymb,amsthm,mathtools}
\usepackage{booktabs,longtable,array}
\usepackage{graphicx}
\usepackage{microtype}
\usepackage{enumitem}
\usepackage{natbib}
\usepackage{xr-hyper}
\usepackage{hyperref}
\usepackage{xcolor}
\usepackage{caption}
\usepackage{subcaption}
\usepackage{float}
\usepackage{verbatim}

\hypersetup{colorlinks=true,linkcolor=blue!55!black,citecolor=blue!55!black,urlcolor=blue!55!black}
\allowdisplaybreaks
\newtheorem{theorem}{Theorem}[section]
\newtheorem{proposition}[theorem]{Proposition}

\theoremstyle{definition}
\newtheorem{assumption}[theorem]{Assumption}

\newcommand{\1}{\mathbf 1}
\newcommand{\E}{\mathbb E}
\newcommand{\Prb}{\mathbb P}
\newcommand{\N}{\mathbb N}

\newcommand{\dd}{\,\mathrm d}

\title{\textbf{Left-Digit Bias: Theory and Empirical Evidence from a Tax Salience Experiment}}
\author{Kaushil Patel\thanks{I am grateful to Rohan Dutta and Larry Epstein for their constant support and generous feedback.}}
\date{November 2, 2026}

\begin{document}
\maketitle
\thispagestyle{plain}

\begin{abstract}
Demand often drops sharply at whole-dollar prices. To explain this phenomenon, known as left-digit bias, we develop a random utility model in which consumers consider a good only if its price is below a heterogeneous attention threshold. Masses of consumers with integer attention thresholds generate demand discontinuities. When a tax is introduced, the model predicts a weaker demand response compared to rational choice for left-digit biased consumers. Using data from a tax experiment, we find that the model matches behavioral moments and partially identify the distribution of willingness to pay relative to thresholds. A weak demand response need not imply a low excess burden of taxation and understanding the specific mechanism is important for welfare analysis. 

%Demand often drops sharply at whole-dollar prices. We develop a random utility model in which consumers consider a good only if its price is below a heterogeneous attention threshold. A mass of consumers with the same attention threshold can generate demand discontinuities even with heterogeneous preferences. When a tax is introduced, the model predicts under-reaction to taxes and a weakly stronger reaction to larger taxes despite consumers fully processing the tax. Using data from a tax salience experiment, we find that the model matches behavioral moments, and we partially identify the distribution of willingness to pay relative to thresholds in the population. Finally, we provide both theoretical and empirical bounds on the excess burden of taxation and compare them to other welfare interpretations. A weak demand response to taxes need not imply a low excess burden.
\end{abstract}

\vspace{0.5em}

\section{Introduction}
\subsection{Overview}
Aggregate demand curves often exhibit sharp discontinuities at whole-dollar prices \citep{ListEtAl2023,StrulovShlain2023}. These discontinuities can be economically large: in the ride-sharing setting studied by \citet{ListEtAl2023}, roughly half of the decline in demand occurs at whole-dollar boundaries. This phenomenon, referred to as \textit{left-digit bias}, has been well-documented in marketing and psychology \citep{CarverPadgett2012,TrollEtAl2024}. This paper presents a model of limited attention that offers a novel explanation for left-digit bias. Using rich data from an experiment studying how consumers respond to sales taxes, we test empirically the model's behavioral implications and partially identify the relationship between consumers' willingness to pay and attention.

%The familiar intuition that inelastic demand implies a small deadweight loss from taxation need not hold when consumer attention depends on price. 

Consider a population of consumers facing a binary choice problem where each consumer decides whether to purchase a consumer good or not. The good has a posted price $p$ and some sales tax may be applied on top of it. For now, assume there is no sales tax, and the posted price is the final amount the consumer pays to purchase the item. After making her choice, each consumer spends her remaining income on a numeraire good. 

Consumers follow a two-stage choice procedure. First, each consumer has a natural-numbered price threshold $T$, which we call an \textit{attention-price threshold (APT)}, and considers a good only if its price is below that threshold. Conditional on consideration, she chooses the good if it maximizes her utility (i.e., her willingness to pay $V$ is greater than $p$).\footnote{To be precise, we allow an infinite attention threshold as well so consumers may consider the good at every price. Also, the empirical application focuses on small ticket items, such as household cleaning products, so we restrict our focus to quasi-linear random utility in the second-stage.} We allow both utility and attention to vary across consumers under the assumption that willingness to pay is atomless. We refer to this model as the \textit{random utility model with attention-price thresholds (RUM--APT)}, and introduce it formally in Section~\ref{sec:model}. 

The choice procedure can be summarized via a posted-price cutoff, $\min\{T,V\}$. A consumer chooses the good if and only if posted price $p<\min\{T,V\}$. We use the term \textit{cutoff} to distinguish it from a consumer's ``true'' willingness to pay $V$. The model produces a demand discontinuity at an integer price if there is a mass of consumers with an attention threshold at that integer and willingness to pay greater than the threshold $T<V$. We refer to these consumers as \textit{binding-threshold consumers}. While the attention threshold is identified by the discontinuity, the demand function does not identify the distribution of willingness to pay above the threshold.

%Consumer demand response to taxes is known to be weaker than implied by the neoclassical model. Prevailing explanations rest on the sales tax not being fully salient since the tax is not visible in the posted price of the good. Can the attention threshold mechanism explain also the demand response to taxes of binding-threshold consumers?

%Next, we ask if the model can explain also the tax behavior of binding-threshold consumers. 
Next, Section~\ref{sec:data} presents the experiment of \citet{TRJ2018} (TRJ), which studies how consumers respond to sales taxes. For each of twenty goods, they ask consumers the following questions: What is the maximum posted price you are willing to pay when there is a sales tax at a given rate? What is the maximum posted price you are willing to pay when there is no sales tax? For the first question, consumers are randomly assigned to face their local sales-tax rate, three times that rate, or no tax. The experiment therefore provides two reported posted-price cutoffs for each consumer and good, with tax variation across experimental groups. In our model, a consumer facing an ad valorem tax at rate $\tau$ chooses the good if $p<\min\{T,V/(1+\tau)\}$; that is, she chooses the good if the posted price is below her attention threshold, and the final amount she has to pay is below her willingness to pay. 

%For each of twenty goods, they ask consumers the following questions sequentially: What is the maximum posted price you are willing to pay when there is a tax equal to your local sales tax? What is the maximum posted price you are willing to pay when there is no sales tax? Thus, the experiment provides two posted price cutoffs. Given our model, a consumer facing an ad valorem tax $\tau$ chooses the good if its posted price is less than the lesser of her attention threshold or her posted price willingness to pay given the tax $V/(1+\tau)$. The experimental set up varies also the sales tax in the first question to be either triple a consumer's local sales tax or zero. Thus, we observe posted price cutoffs with rich tax variation.

The RUM--APT model makes sharp predictions about these cutoffs, as derived in Section~\ref{sec:hybrid_tax}. Every binding-threshold consumer under-reacts to taxes, where under-reaction is defined as a demand response weaker than that implied by rational choice. Under-reaction to taxes weakly decreases as the tax rate rises. We refer to a consumer's cutoff moving below her binding threshold as \textit{debunching}. The model predicts that the mass remaining at binding thresholds weakly decreases as the tax rate rises.

%Every binding-threshold consumer must under-react to the tax but under-react weakly less as the tax rate rises, where under-reaction is defined as a demand response weaker than that implied by rational choice, and the mass remaining at binding thresholds weakly decreases in the tax rate (we refer to a consumer's cutoff moving from a binding threshold as \textit{debunching}). 

In Section~\ref{sec:identification} , we ask what can be learned about consumer preferences from cutoff data. Tax response and debunching are governed by the same distribution of willingness to pay relative to the attention threshold, $V/T$, which we refer to as the \textit{valuation overshoot}.  With population moments at finitely many tax rates, debunching identifies values of the overshoot CDF, while average tax response identifies integrals of that CDF. Thus, population moments allow partial identification of the overshoot distribution. If we had noiseless individual cutoff data, then $V$ would be point-identified for any binding-threshold consumer whose cutoff changes under the tax.

Sections~\ref{sec:estimation} and~\ref{sec:results} present our estimation strategy and empirical results, respectively. For our empirical application, we assume that integer cutoff reports correspond to binding-threshold observations. Such an assumption is motivated by the existence of large spikes in the number of cutoffs at integer prices. These account for around 13 percent of consumer-product cutoff reports with no sales tax. The estimated mean tax response among binding-threshold consumers is 0.27 with the local sales tax and 0.38 with triple the local sales tax. Intuitively, average tax response to the local sales tax is about a quarter of that expected from a rational decision maker. Estimated debunching--share of binding-threshold consumers whose cutoff changes due to the tax--is 41.5 percent and 52.2 percent, respectively. Both comparisons are also directionally consistent with stronger responses to larger taxes, although neither difference between the two tax environments is statistically significant. The tax responses for remaining consumers are 0.247 and 0.488, similar to the estimates from \citet{TRJ2018}.

%Reports differ substantially between modules even without taxation, so we use the no-tax arm to control for such noise when estimating tax response and debunching in the taxed arms. 

%The estimated mean tax response among binding-threshold consumers is 0.269 in the standard-tax arm and 0.383 in the triple-tax arm. Intuitively, average consumer response in the standard arm is about a quarter of that expected of a rational decision maker. Estimated debunching is 41.5 percent and 52.2 percent, respectively. Both comparisons are also directionally consistent with stronger responses to larger taxes, although neither cross-arm difference is statistically significant. The tax responses for remaining consumers are 0.247 and 0.488, closely replicating the estimates from \citet{TRJ2018}.

Next, we ask whether one overshoot distribution can explain the two tax response moments and the two debunching moments jointly. Our specification reproduces all four estimated moments. Considering all distributions that satisfy these restrictions gives informative bounds on the overshoot distribution: for example, between 43.8 and 52.2 percent of binding-threshold consumers have willingness to pay no more than 10 percent above their threshold.

Finally, in Section~\ref{sec:welfare} we conduct welfare analysis. Any pair of no-tax and taxed cutoffs for a binding-threshold consumer that can be rationalized by RUM--APT can also be explained by TRJ's framework, but their welfare implications differ. TRJ interpret the no-tax cutoff as willingness to pay. Given this assumption, their welfare claims do not rest on a particular mechanism driving the consumer's tax response. Guided by our model, we interpret a binding cutoff---in any tax environment---as an attention threshold below willingness to pay. A taxed cutoff below the threshold then reveals a consumer's willingness to pay. Under perfectly elastic supply, the same purchase eliminated by taxation entails a larger surplus loss under RUM--APT. These results illustrate why a weak demand response need not imply a small welfare cost: the lost surplus from trade depends on the behavioral mechanism.

%Following the robust behavioral welfare analysis approach pioneered by \citet{BernheimRangel2009} \textbf{(and followed by recent papers in behavioral public economics)}, TRJ interpret certain choices as optimal and relevant for welfare analysis, while treating others as suspect and not reflecting the consumer's true willingness to pay. In the tax experiment, they interpret the no-tax cutoff as willingness to pay. Given this assumption, their welfare claims do not rest on a particular mechanism driving the consumer's tax response. 

%Our approach can be described as behavioral revealed preference analysis.\footnote{\textbf{Differences in the two approaches have been discussed at length in the axiomatic choice theory literature (see)}.} Guided by our model, we interpret a binding cutoff---in any tax environment---as an attention threshold below willingness to pay. A taxed cutoff below the threshold then reveals a consumer's willingness to pay. The same purchase eliminated by taxation entails a larger surplus loss under RUM--APT. These results illustrate why a weak demand response need not imply a small welfare cost: the lost surplus from trade depends on the behavioral mechanism; RUM--APT can imply greater excess burden than even the neoclassical model. 

For a given price, we derive bounds on the average excess burden of taxation under the two models, and their ratio. The upper bound on the ratio grows as the price approaches an integer from below. Recent work has found that firm pricing fails to fully account for left-digit bias among consumers \citep{ListEtAl2023,StrulovShlain2023}. If firms optimize by raising prices to ``just below'' an integer, then the two models can imply increasingly different welfare costs relative to one another.  

%To obtain quantitative welfare bounds, we additionally assume that the overshoot distribution is common across attention thresholds. This connects the aggregate evidence about valuations relative to thresholds to purchase decisions at a given price. At $\$4.95$ with a 10 percent tax and perfectly elastic supply, the estimated moment restrictions imply a ratio of mean excess burdens of 3.15--4.73, compared to theoretical bounds of 1-9.9. 

While our analysis has focused on binding-threshold consumers, the economic significance of our welfare results may be larger than suggested by the share of binding-threshold observations in \citet{TRJ2018}. Using the experimental data from \citet{CLK2009}, \citet{Ashton2014} provides reduced-form evidence that if the posted price of an item includes the tax, then the demand response is strong only when the tax inclusion causes the posted price to cross an integer. Thus, left-digit bias may be a large channel for under-reaction to taxes. Section~\ref{sec:discussion} discusses the connection to the literature on left-digit bias and behavioral welfare analysis.

%Section~\ref{sec:data} presents the experiment and the TRJ framework. Sections~\ref{sec:hybrid_tax} and~\ref{sec:identification} derive behavioral implications and identification results. Section~\ref{sec:estimation} describes estimation, Section~\ref{sec:results} presents the empirical results, and Section~\ref{sec:welfare} develops the welfare comparison. Section~\ref{sec:discussion} discusses connections to left-digit bias and behavioral welfare analysis. The appendices contain the proofs and estimation details.

\subsection{Related Literature}

The decision model studied here is related to the literature on consideration sets and two-stage choice procedures \citep{HauserWernerfelt1990,MasatliogluEtAl2012,ManziniMariotti2007,ManziniMariotti2012,ManziniMariotti2014a}. In our model, consumers apply a price threshold to form a consideration set, a subset of the set of feasible alternatives. Then, in the second stage, they choose their most preferred alternative from the consideration set. The literature explores various ways consideration sets can be formed. 

Two papers that focus on attention thresholds to form consideration sets are \citet{JagabathulaRusmevichientong2017} and \citet{Patel2025}. \citet{JagabathulaRusmevichientong2017} use the RUM--APT model, while allowing attention thresholds at any price, to study the firm's joint assortment and price choice problem with multiple goods; they do not analyze the behavioral content of the model or conduct consumer welfare analysis. \citet{Patel2025} studies the binary choice model with a common preference across consumers in the second stage; he provides a behavioral characterization and identification of welfare change when prices change. We add to the study of the attention-price threshold mechanism by demonstrating the ability of the model to explain left-digit bias and testing empirically its predictions for consumers' responses to taxes. In addition, our welfare analysis relates to the behavioral public economics literature on the welfare consequences of taxation. 

%Our work contributes also to the discrete choice literature on consumer demand when there is heterogeneity in both preferences and attention. With both sources of heterogeneity, identification of the distribution of preferences, and consequently welfare analysis, is generally difficult (\textbf{see Crawford et al. 2021 for a review}). Our approach in this paper shows how empirical observations about other behavioral anomalies, e.g., left-digit bias, can guide reasonable restrictions on the role of attention and preference (e.g., natural-numbered attention thresholds) and permit welfare analysis. 

%\textbf{Abaluck and Adams (2021)} provide a positive result where for some models of random attention and quasi-linear random utility, identification is possible from choice data when there are at least three goods. \citet{Patel2025} points out that for binary choice the attention threshold model with a homogeneous preference can rationalize any choice behavior that the quasi-linear random utility model can.

%The paper contributes towards a theory of consumer demand by parsimoniously extending the random utility model to incorporate left-digit bias and exploring its implications for consumer welfare. 
In behavioral public economics, the ``model-free'' welfare analysis approach of \citet{BernheimRangel2009} has been used extensively (see \citet{BernheimTaubinsky2018} for a review). The broad idea is to specify which choices are informative about welfare and then to infer welfare comparisons from those choices without relying on a specific model of choice. TRJ follow this approach by assuming that posted-price cutoff reports when there is no sales tax are rational and reflect the consumer's willingness to pay. Our welfare analysis shows that under RUM--APT the excess burden of taxation can differ meaningfully from that under the TRJ interpretation, and the specific mechanism generating the under-reaction to taxes matters. The differences between model-free and model-based approaches to behavioral welfare analysis have been discussed in the choice theory literature \citep{MasatliogluEtAl2012,RubinsteinSalant2012,ManziniMariotti2014Welfare}.

%Following the robust behavioral welfare analysis approach pioneered by \citet{BernheimRangel2009} \textbf{(and followed by recent papers in behavioral public economics)}, TRJ interpret certain choices as optimal and relevant for welfare analysis, while treating others as suspect and not reflecting the consumer's true willingness to pay. In the tax experiment, they interpret the no-tax cutoff as willingness to pay. Given this assumption, their welfare claims do not rest on a particular mechanism driving the consumer's tax response. 

%Our approach can be described as behavioral revealed preference analysis.\footnote{\textbf{Differences in the two approaches have been discussed at length in the axiomatic choice theory literature (see)}.} Guided by our model, we interpret a binding cutoff---in any tax environment---as an attention threshold below willingness to pay. A taxed cutoff below the threshold then reveals a consumer's willingness to pay. The same purchase eliminated by taxation entails a larger surplus loss under RUM--APT. These results illustrate why a weak demand response need not imply a small welfare cost: the lost surplus from trade depends on the behavioral mechanism; RUM--APT can imply greater excess burden than even the neoclassical model. 

Our work contributes a new explanation for left-digit bias. Previous explanations have viewed the behavior as arising from misperception of prices ending in 99 cents as being lower than they objectively are \citep{DellaVigna2009,LaceteraPopeSydnor2012}. There is mixed evidence for such a mechanism from experiments on price recall, perceived price differences, and purchasing behavior \citep{SchindlerKibarian1996,ManningSprott2009,CarverPadgett2012,TrollEtAl2024}. Our model does not assume consumers misperceive the price. We test our model in the context of a sales tax experiment. A distinct feature of the tax experiment is that consumers are not presented with a product at a given price and then asked if they would buy, but rather are presented with a product and asked to report the maximum posted price they are willing to pay. 

%The paper contributes towards a theory of consumer demand by parsimoniously extending the random utility model to incorporate left-digit bias and exploring its implications for consumer welfare. 

We contribute also to the literature on tax salience and the role of ``not-fully-salient'' incentives in decision making \citep{GabaixLaibson2006,Finkelstein2009,FeldmanRuffle2015,FeldmanGoldinHomonoff2018,KroftEtAl2024}. Salience refers to the extent to which consumers account for an incentive in their decision-making. Previous explanations for under-reaction to taxes have focused on a tax as being a special attribute of the decision problem that may not be fully salient to the consumer \citep{TRJ2018,CLK2009}. In this paper, we show that even with complete processing of the tax, consumers following the heuristic of an attention threshold will under-react to taxes. An important question has been whether salience may be endogenous to the strength of the incentive. For example, are higher taxes more salient? TRJ accommodate such endogeneity by permitting their salience parameter to vary with the tax rate. We show that even a fixed attention threshold can produce the pattern of weaker under-reaction to higher taxes. As discussed earlier, distinguishing between these explanations is important for welfare analysis.

While we focus on left-digit bias and binding-threshold consumers, TRJ's main finding is that consumers overwhelmingly under-react to taxes, including those with non-integer cutoffs. If we allow attention-price thresholds to be any positive real number, then the model can explain under-reaction to taxes for those consumers as well.\footnote{\citet{MorrisonTaubinsky2023} find that while there is consistent evidence of under-reaction on average, some consumers do over-react to taxes. Over-reaction cannot be explained by our model with fixed attention thresholds.}

%one may ask whether the attention threshold mechanism can explain the under-reaction of all the consumers in the TRJ experiment. The answer is yes. If we allow attention-price thresholds to be any positive real number, then the model can explain under-reaction to taxes for the full population in the experiment, including those with non-integer cutoffs.\footnote{There is recent evidence from \citet{MorrisonTaubinsky2023} that while at the population level, there is consistent evidence of strong under-reaction, some consumers do overreact to taxes, which cannot be explained by our model with fixed attention thresholds.} 

\section{RUM--APT and ordinary cross-sectional demand}\label{sec:model}

\subsection{Preferences}

There are two alternatives: $0$ and $1$. The price of good $0$ is zero, and the price of good $1$ is $p\geq0$. Good $1$ is a consumer good, and good $0$ represents the outside option of not choosing the consumer good. There exists a unit mass of consumers. Each consumer has income $y>0$, which is common across consumers. She picks one of the two discrete goods and spends her residual income on a numeraire good. The quantity of numeraire consumed when good $0$ or good $1$ is chosen is $y$ and $y-p>0$, respectively.

We have in mind consumption goods for which prices are small relative to income, such as cereals or household cleaning products. In this setting, income effects are likely negligible. We assume that consumer preferences are described well by a quasi-linear random utility model. The utilities for good $0$ and good $1$ are
\[
u_{0}(y)=y,\qquad u_{1}(y-p)=y-p+V,
\]
where $0\leq V<y$ is the willingness to pay for good $1$, which can vary across consumers. Good $1$ is preferred to good $0$ if $V>p$. Thus, one source of choice heterogeneity is preference heterogeneity.

\begin{assumption}[No utility ties]\label{assnoties}
For every $p\geq0$, $\Pr(V=p)=0$.
\end{assumption}

The no utility ties assumption implies that the distribution of $V$ has no atoms.

\subsection{Natural attention thresholds}

Decision makers follow a two-step procedure to choose an alternative. In the first step, the consumer considers those alternatives with prices less than $T$. We call $T$ the \emph{attention-price threshold} for the consumer, and the set of alternatives she considers her consideration set. In the second step, she picks the utility-maximizing alternative from her consideration set. Importantly, different consumers may have different attention-price thresholds.

\begin{assumption}[Natural attention thresholds]\label{assthresholdsupport}
The attention-price threshold satisfies
\[
T\in\mathbb{N}_{+}\cup\{\infty\}
=\{1,2,\ldots\}\cup\{\infty\}.
\]
\end{assumption}

The support of consumers' attention thresholds is restricted to the natural numbers and infinity. A consumer with an infinite threshold pays attention to good $1$ at all prices. A consequence of the support restriction is that good $0$ is considered by all consumers.

Combining our preference and attention structure, each consumer is described by a pair $(V,T)$. Thus, the model permits both preference and attention heterogeneity. Beyond Assumptions~\ref{assnoties} and~\ref{assthresholdsupport}, the joint distribution of $(V,T)$ is unrestricted. In particular, preference and attention heterogeneity may be arbitrarily correlated.

Define the effective price cutoff
\begin{equation}\label{eqeffectivecutoff}
C:=\min\{T,V\}.
\end{equation}

A consumer chooses good $1$ if $p<C$. The price cutoff $C$ gives the consumer's effective willingness to pay for the good. Given limited attention, this may not equal the consumer's "true" willingness to pay, $V$, which is the price cutoff she would act upon if she was rational, i.e., fully attentive. The choice probability for good $1$ is
\begin{equation}\label{eqdemand}
Q(p)=\Pr(C>p).
\end{equation}
Since there is a unit mass of consumers, $Q(p)$ is also aggregate demand for good $1$.

We refer to this model as the RUM--APT model. Preference heterogeneity is captured by the quasi-linear random utility model and attention heterogeneity is captured by the attention-price threshold (APT) model of \citet{Patel2025}. Unlike the quasi-linear random utility model with atomless willingness to pay, RUM--APT permits discontinuous choice behavior, and unlike the APT model, it permits preference heterogeneity.\footnote{Another difference from the APT model of \citet{Patel2025} is that the support of attention thresholds is restricted to the natural numbers, whereas \citet{Patel2025} permits the support to be the set of nonnegative real numbers.}

\subsection{Implications for demand}\label{seccrosssectionimplications}

Consider what the model implies for aggregate demand, $Q(p)$. This is the empirical object one would estimate from standard cross-sectional consumer choice data. Because individual choice depends only on the effective price cutoff $C=\min\{T,V\}$, the choice-probability function identifies the distribution of $C$. Generally, it is not possible to identify the joint distribution of $(V,T)$. However, the positive-integer support restriction on attention thresholds allows us to identify some features of the distribution of attention.\\

\textbf{Observation 1 (Integer atoms).} \emph{With atomless willingness to pay, any discontinuity in the choice probability, $Q(p)$, must occur at a positive integer.}\\

% To see this, fix any noninteger $x>0$. Because $T$ takes values
% only in $\mathbb{N}_{+}\cup\{\infty\}$, the event $\{C=x\}$ can
% occur only if $V=x$. Assumption~\ref{assnoties} therefore implies
% \[
% \Pr(C=x)=0
% \qquad
% \text{for every }x>0\text{ with }x\notin\mathbb{N}_{+}.
% \]
%
% At an integer $n\in\mathbb{N}_{+}$, by contrast, a positive atom
% can arise because the attention threshold binds.

For $n\in\mathbb{N}_{+}$, define the size of the downward discontinuity in the choice probability at $n$ by
\[
J_n:=Q(n-)-Q(n).
\]
Because $Q(p)=\Pr(C>p)$,
\begin{equation}\label{eqJidentity}
J_n=\Pr(C=n)=\Pr(T=n,V>n).
\end{equation}
Thus, an integer discontinuity identifies the mass of consumers whose attention threshold is exactly $n$ and whose willingness to pay lies above that threshold. It does not identify the total mass of consumers with threshold n, $\Pr(T=n)$.

\begin{proposition}[Identification from cross-sectional demand]\label{propnotaxid}

Suppose we observe the choice probability function $Q(p)$. The discontinuity $J_n$ identifies the mass of consumers with a binding threshold at $n$: $\Pr(T=n,V>n)=J_n$.
\end{proposition}

Cross-sectional data does not identify the total mass of consumers with finite attention thresholds. A consumer with $V<T<\infty$ can instead be assigned $T=\infty$ without changing her effective price cutoff: in either case $C=V$. We also do not learn about the distribution of $V$ above $n$ for consumers with binding thresholds. Holding the binding mass at $n$ and all other consumers unchanged, any atomless conditional distribution of $V$ on $(n,y)$ will produce the same choice probability function. Thus, ordinary cross-sectional demand identifies the binding integer masses, but neither the total finite-threshold population nor the distribution of willingness to pay above a binding threshold.

\paragraph{Connection to left-digit bias.}

There is robust evidence for demand discontinuities at whole-dollar prices. \citet{ListEtAl2023}, for example, find large demand discontinuities at dollar values in a field experiment at Lyft, while \citet{StrulovShlain2023} emphasizes a ``sawtooth'' demand pattern in which demand is comparatively flat within a dollar interval and drops discretely when the dollar amount changes. The behavioral economics literature interprets these demand discontinuities as \emph{left-digit bias}. The RUM--APT model provides a novel explanation for these demand discontinuities. See Section~\ref{sec:discussion} for more discussion. 

%The canonical empirical signature of left-digit bias is a disproportionate change in demand when a price crosses a whole-dollar boundary.

\section{The Taubinsky and Rees-Jones (2018) tax salience experiment}\label{sec:data}

Standard choice data, $Q(p)$, provides a testable restriction for attention thresholds (demand discontinuities at integer prices) and reveals some interesting features (the mass of consumers with binding attention thresholds). However, this is a weak test and an important primitive of the model, the distribution of $V$ among consumers with binding attention thresholds, remains unidentified. To obtain more demanding tests of the attention threshold mechanism and sharper identification, we turn to a setting with richer variation: the tax experiment of \citet{TRJ2018}.

\subsection{Experimental design}

The experiment involved 4,223 consumers from the United States, each making decisions for 20 household products. The experiment can be divided into three arms and two modules within each arm. Each consumer was randomly assigned to a no-tax, standard-tax, or triple-tax arm. The differences across the three arms arise in Module 1. In Module 1 of the no-tax arm, consumers were asked to report the highest tag price they were willing to pay for each of the twenty household products when there was no tax. In Module 1 of the standard-tax arm, consumers were asked to report the highest tag price they were willing to pay for those products when the tax was equal to their local sales tax. In Module 1 of the triple-tax arm, consumers were asked to report the highest tag price they were willing to pay when the tax was three times their local sales tax.

After completing Module 1, consumers faced Module 2. Module 2 was identical across all three arms and asked consumers to report the highest tag price they were willing to pay for the same 20 products when there was no tax. Thus, for each consumer-product pair, we observe a Module-1 cutoff and an untaxed Module-2 cutoff. Consumers in the no-tax arm faced the same choice problem in both modules (the same products and no tax), providing a control group for Module-1-versus-Module-2 order effects and reporting differences.

The willingness-to-pay reports were incentivized using the Becker-DeGroot-Marschak (BDM) mechanism. A consumer purchased the product at a randomly selected price if that price was below her reported cutoff. Consumers were given an explanation of the BDM mechanism and informed that it was in their interest to report honestly the highest tag price they were willing to pay.

Let $i$ index consumers and $k$ index products. Let $P^1_{ik}$ and $P^2_{ik}$ denote consumer $i$'s reported cutoff for product $k$ in Module 1 and Module 2, respectively. We refer to these experimental responses as reported cutoffs to distinguish them from the consumer's ``true'' willingness to pay, $V$, in the RUM--APT model. For consumers in the standard-tax and triple-tax arms, $P^1_{ik}$ is elicited under a positive sales tax while $P^2_{ik}$ is elicited without a tax. For consumers in the no-tax arm, both $P^1_{ik}$ and $P^2_{ik}$ are elicited without a tax.

\subsection{The Taubinsky and Rees-Jones tax-salience framework}\label{sec:trj_benchmark}

\citet{TRJ2018} model imperfect response to sales taxes by introducing a new parameter to govern the tax response. Let $V$ denote the consumer's ``true'' willingness to pay, let $\tau\geq0$ denote the ad valorem sales-tax rate, and let $\theta\geq0$ denote how much the consumer under- (or over-) reacts to the tax. Under quasi-linear utility, the highest tag price at which the consumer is willing to purchase the product is

\begin{equation}\label{eq:smoothcutoff}
C_{\mathrm{TRJ}}(\tau)=\frac{V}{1+\theta\tau}.
\end{equation}

When $\theta=0$, the consumer ignores the tax completely; when $\theta=1$, she responds fully to the tax; and when $\theta>1$ she over-reacts to the tax.

Importantly, $\theta$ is defined behaviorally. For a consumer-product pair with $\tau>0$, the model defines $\theta$ as
\[
\theta=\frac{P^2/P^1-1}{\tau},
\]
where $P^2$ and $P^1$ are the consumer's reported cutoffs with no tax and with tax rate $\tau$, respectively. $\theta$ can be interpreted as inattention to the tax, but given its behavioral definition, it also accommodates various other explanations for imperfect tax response, as described in \citet{TRJ2018}.

\citet{TRJ2018} maintain the quasi-linear utility assumption for both their theoretical and empirical analysis. The support of $\theta$ is restricted to be nonnegative and bounded. The distribution of $\theta$ can depend on the tax, so attention to taxes can depend on the tax rate. However, the marginal distribution of $V$ is assumed not to depend on the tax. They assume that the joint distribution of $V$ and $\theta$ generates smooth and downward-sloping aggregate demand curves. Otherwise, the joint distribution of $V$ and $\theta$ is allowed to be flexible.

The main empirical finding of \citet{TRJ2018} is substantial under-reaction to sales taxes. The estimated average tax response is $0.250$ in the standard-tax arm and $0.475$ in the triple-tax arm. Consumers respond on average to only about one-quarter of the standard tax and about one-half of the triple tax. In short, consumers under-react to taxes, but they are more attentive to larger taxes.

\section{RUM--APT with tax variation: behavioral implications}\label{sec:hybrid_tax}

Under-reaction to taxes cannot be explained by the random utility model. Consumers, modeled by RUM, behave rationally and react fully to the tax. Thus, we will begin by modifying the RUM--APT model, specifically the RUM component of the model. We will allow fully attentive consumers (with $T=\infty$) to behave as in the model of Taubinsky and Rees-Jones (2018) from Section 3.2 (henceforth, referred to as TRJ). We make no changes to consumers with finite attention thresholds. What behavioral implications do attention thresholds have for the TRJ experiment?

\subsection{Tax behavior and the continuous residual}\label{sec:hybrid_specialization}

Recall from Section~2 that each consumer is characterized by willingness to pay $V$ and an attention threshold $T\in\mathbb N_+\cup\{\infty\}$. Her no-tax posted-price cutoff is $C=\min\{T,V\}$. We now specify how consumers behave when the product is subject to an ad valorem tax $\tau\geq0$. 

We allow consumers with $T=\infty$ to follow the TRJ framework. Their posted-price cutoff is
\[
C_{\mathrm{TRJ}}(\tau)=\frac{V}{1+\theta\tau},
\]
where $\theta\geq0$ is the consumer's behavioral tax-response parameter. With this change, these consumers can now under- and over-react to taxes.

Next, we turn to consumers with finite attention thresholds. Attention continues to depend on the posted tag price $p$: the product is considered if $p<T$. Conditional on considering the product, however, the consumer fully processes the tax. She prefers the product to the outside option if $p<\frac{V}{1+\tau}$. Therefore, her posted-price cutoff under tax $\tau$ is 
\begin{equation}\label{eq:thresholdcutoff}
C_A(\tau)=\min\left\{T,\frac{V}{1+\tau}\right\}.
\end{equation}
At $\tau=0$, equation~\eqref{eq:thresholdcutoff} reduces to the no-tax cutoff from Section~2: $C_A(0)=C=\min\{T,V\}$.

Consumers with finite attention thresholds can be divided into two groups: those with $V<T$ and those with $V\geq T$. A finite threshold consumer with $V<T$ is observationally equivalent in the experiment to a consumer with $T=\infty$ and $\theta=1$ because the attention threshold never binds. For these finite-threshold types, we set $\theta=1$. We refer to the union of $T=\infty$ tax-salience types and finite-threshold types with $V<T$ as the \emph{continuous residual}. With no taxes, these consumers do not contribute to demand discontinuities. Every continuous-residual type can therefore be written as having posted-price cutoff
\begin{equation}\label{eq:continuouscutoff}
C_{\mathrm{cont}}(\tau)=\frac{V}{1+\theta\tau}.
\end{equation}

\begin{assumption}[Atomless continuous residual]\label{ass:continuousnoatoms}
For every tax rate $\tau$, the distribution of the continuous residual posted-price cutoff $C_{\mathrm{cont}}(\tau)$ is atomless.
\end{assumption}

Assumption~\ref{ass:continuousnoatoms} ensures that the continuous residual types do not generate any atoms in the cutoff distribution. It rules out joint distributions of $V$ and $\theta$ that can generate atoms in response to taxation. 

The remaining distinct component consists of binding finite-threshold types with $T=n<V$ for some natural number $n$. These are consumers whose no-tax cutoff is determined by their attention threshold rather than by their willingness to pay, and they can generate demand discontinuities in the no-tax choice probability function, $Q(p)$. We make no modifications to them, and their posted-price cutoffs are given by equation~\eqref{eq:thresholdcutoff}. The behavior of these consumers is our primary concern. Specifically, we want to know if attention-price thresholds can explain their behavior in the tax salience experiment. 

\subsection{Tax-induced debunching and mover support}

Consider a binding finite-threshold type with $T=n<V$. Her no-tax cutoff is $C=n$. Under a positive tax rate $\tau$, her cutoff remains at $n$ if and only if
\[
V\geq n(1+\tau).
\]
Otherwise, the attention threshold ceases to bind, and her cutoff falls below $n$. Because the continuous residual is atomless by Assumption~\ref{ass:continuousnoatoms}, any atom at $n$ in the cutoff distribution under tax must consist entirely of surviving binding threshold types. Extending the notation from Section~2, let $Q_\tau(p)$ denote the choice probability when the posted price is $p$ and the tax rate is $\tau$, and let the demand discontinuity at integer $n$ be
\begin{equation}\label{eq:Jtax}
J_n(\tau)
:=
Q_\tau(n-)-Q_\tau(n).
\end{equation}
At zero tax, $J_n(0)$ is the demand discontinuity defined in Section~2. Under the model,
\begin{equation}\label{eq:Jtax_structural}
J_n(\tau)
=
\Prb\!\left(T=n,\;V>n(1+\tau)\right).
\end{equation}
The distinction between $>$ and $\geq$ is immaterial by Assumption~\ref{assnoties}.

\begin{proposition}[Tax-induced debunching]\label{prop:debunch} 
We refer to a tax-induced cutoff change from a binding attention threshold as debunching.
\begin{enumerate}
    \item  For every positive integer price $n$, the demand discontinuity, $J_n(\tau)$, is weakly decreasing in the tax rate. 
    \item Whenever $J_n(0)>0$, the share of debunching consumers is given by
\begin{equation}\label{eq:debunchcdf}
1-\frac{J_n(\tau)}{J_n(0)}
=
\Prb\!\left(
\frac{V}{T}\leq1+\tau
\,\middle|\,
T=n,V>n
\right).
\end{equation}
\end{enumerate}
\end{proposition}

Proposition~\ref{prop:debunch} provides the first sharp implication of attention thresholds with tax variation. A tax can only reduce the mass at a binding integer threshold, and a larger tax weakly reduces more of that mass. The amount of debunching reveals information about willingness to pay above the threshold that ordinary no-tax demand does not identify. Among consumers whose no-tax cutoff is determined by $T=n$, the fraction that leaves the atom under tax $\tau$ is the fraction whose willingness to pay is no greater than $1+\tau$ times their threshold.

The model also restricts where the mass that leaves an integer can go. Once the attention threshold ceases to bind, a threshold-$n$ consumer's taxed cutoff is $V/(1+\tau)$. Assumption~\ref{assnoties} implies that the distribution of this cutoff among movers is atomless. Together with Assumption~\ref{ass:continuousnoatoms}, this implies that taxation cannot generate a new atom elsewhere in the cutoff distribution.

\begin{proposition}[No new atoms under taxation and mover support]\label{prop:nonewatoms}
Let $\tau>0$. 
\begin{enumerate}
    \item The taxed cutoff distribution has no atoms at noninteger prices.
    \item The taxed cutoff for a binding threshold-$n$ type who moves in response to the tax satisfies
\begin{equation}\label{eq:moversupport}
\frac{n}{1+\tau}
<
C_A(\tau)
<
n
,
\end{equation}
and the conditional distribution of $C_A(\tau)$ among these movers is atomless.
\end{enumerate}
\end{proposition}

Proposition~\ref{prop:nonewatoms} gives two additional restrictions. First, taxation can shrink or eliminate an existing integer atom but cannot relocate that mass into an atom at another price. Second, a consumer who moves away from a binding threshold $n$ must have a taxed cutoff inside
\[
\left(\frac{n}{1+\tau},\,n\right).
\]
Thus the model predicts both \emph{atom destruction without atom relocation} and a threshold-specific support restriction for movers.

\subsection{Threshold response and the valuation overshoot}\label{sec:overshoot_behavior}

Tax-induced debunching reveals whether willingness to pay lies sufficiently close to a binding attention threshold for the tax to make the threshold cease to bind. The magnitude of the threshold response contains additional information about how far willingness to pay lies above that threshold. Both outcomes are governed by the valuation overshoot
\begin{equation}\label{eq:R}
R:=\frac{V}{T}>1,
\end{equation}
which measures willingness to pay relative to the attention threshold.

If $R$ is close to one, willingness to pay lies only slightly above the threshold, so even a relatively small tax can make the threshold cease to bind. If $R$ is large, willingness to pay lies farther above the threshold and the integer cutoff survives a larger tax.

We define the tax response using the same behavioral definition as in \citet{TRJ2018}. For $\tau>0$, define the tax response of a binding threshold type by
\begin{equation}\label{eq:psiA}
\psi_A(\tau)
:=
\frac{C_A(0)/C_A(\tau)-1}{\tau}.
\end{equation}
A full tax response corresponds to $\psi_A(\tau)=1$. Note that all objects on the right-hand side of the definition are observable. For a TRJ consumer,
\[
\frac{C_{\mathrm{TRJ}}(0)/C_{\mathrm{TRJ}}(\tau)-1}{\tau}
=\theta.
\]
The next proposition states the implications of the attention threshold model for the tax response of binding-threshold consumers.
\begin{proposition}[Threshold tax response]\label{prop:thresholdresponse}
For every binding threshold type and every $\tau>0$, the model implies
\begin{equation}\label{eq:psiA_formula}
\psi_A(\tau)
=\left\{
\begin{array}
[c]{ccc}%
0 &  & \text{if} \ R\ge1+\tau\\
\frac{1+\tau-R}{R\tau} &  & \text{if} \ R< 1+\tau\\
\end{array}
\right.
\end{equation}
Hence, $0\leq\psi_A(\tau)<1$. The tax response $\psi_A(\tau)$ is weakly increasing in $\tau$, and it is strictly increasing once $\tau>R-1$. Consequently, for any fixed distribution of $R$ in a binding-threshold population and $0<\tau_1<\tau_2$,
\[
\E[\psi_A(\tau_1)]
\leq
\E[\psi_A(\tau_2)].
\]
\end{proposition}

Binding threshold consumers under-react to taxes, but their response weakly increases with the size of the tax. At low tax rates, the attention threshold can absorb the entire tax change and leave the cutoff unchanged. Once the tax is large enough for $V/(1+\tau)$ to fall below $T$, the threshold ceases to bind and subsequent tax increases are fully incorporated into the cutoff. Thus, the overall tax response, $\psi_A$, remains below one because it includes the initial interval of inaction, but it rises once the threshold ceases to bind. An interesting prediction (which we will not be able to test given our data) is that every binding-threshold consumer is completely unresponsive to some low enough tax rate. 

In the TRJ model, the same tax response equals $\theta$, which can be interpreted as their attention parameter. The distribution of $\theta$ is allowed to depend on the tax rate, so their model imposes no corresponding cross-tax monotonicity restriction. In our model, the attention threshold is static and does not change with the tax. Monotonicity in the response to taxes is an implication of the choice procedure even with a fixed attention threshold. 

For our empirical analysis, it is useful to work with log cutoff changes. Define the log threshold response of a binding threshold type, for $\tau>0$, as
\begin{equation}\label{eq:phib}
\phi_A(\tau)
:=
\frac{\log C_A(0)-\log C_A(\tau)}{\tau}.
\end{equation}
Let $[x]_+:=\max\{x,0\}$ for the positive part of a real number $x$. Since $C_A(0)=T$, equation~\eqref{eq:thresholdcutoff} implies
\begin{equation}\label{eq:phib_formula}
\phi_A(\tau)
=
\frac{[\log(1+\tau)-\log R]_+}{\tau},
\end{equation}
where $\phi_A$ measures the log cutoff change per unit of the tax rate. We refer to this quantity as log threshold response throughout.

The log response measure should be distinguished from the behavioral response in equation~\eqref{eq:psiA}. Since $\log x\leq x-1$ for every $x>0$,
\begin{equation}\label{eq:phi_psi_order}
0\leq\phi_A(\tau)\leq\psi_A(\tau)<1,
\end{equation}
with $\phi_A(\tau)<\psi_A(\tau)$ whenever the cutoff moves. For a fixed threshold type, the gap $\psi_A(\tau)-\phi_A(\tau)$ is weakly increasing in the tax rate and strictly increasing once the cutoff moves. Thus, the log response measure compresses behavioral response more strongly at larger tax rates. Unlike $\psi_A$, the log tax response need not be increasing in the tax rate. This non-monotonicity comes from using log changes per unit of tax rather than from weaker behavioral responsiveness. Even a fully responsive consumer generates a log response of $\log(1+\tau)/\tau$, which is decreasing in the tax rate $\tau$. We will keep these differences in mind when evaluating our log response estimates. 

Equation~\eqref{eq:phib_formula} also shows the connection between log threshold response and debunching. If $R\geq1+\tau$, the attention threshold continues to bind, so the consumer remains at the integer cutoff and $\phi_A(\tau)=0$. If $R<1+\tau$, then the threshold ceases to bind and the consumer leaves the integer atom. 

To state the resulting restrictions on the overshoot distribution, consider a population of binding-threshold consumers with overshoot $R=V/T$. Let
\[
Z:=\log R>0
\]
and denote its CDF by $F_Z$. For a given tax rate $\tau>0$, let
\[
a:=\log(1+\tau).
\]
Then,
\begin{equation}\label{eq:stoploss}
\tau\,\E[\phi_A(\tau)]
=
\E[(a-Z)_+]
=
\int_0^a F_Z(z)\,\dd z.
\end{equation}
The integral has a simple interpretation. A binding-threshold consumer debunches if $Z<a$. For such a consumer, $a-Z$ measures how far her log overshoot lies below the value required for the attention threshold to continue binding. Consumers with $Z\geq a$ do not debunch and contribute zero to the tax response. Whenever the debunching probability $\Prb(Z<a)>0$,
\[
\E[(a-Z)_+]
=
\Prb(Z<a)\,\E[a-Z\mid Z<a].
\]
Thus, the average log threshold response reflects both the share of binding-threshold consumers who debunch and the magnitude of their responses. Two overshoot distributions can generate the same debunching probability but different average log tax responses. If the log overshoots of consumers who debunch are concentrated close to $a$, their log threshold responses will be relatively small. If their log overshoots lie farther below $a$ (and thus their valuations lie closer to their attention threshold), then their log threshold responses will be larger.

At the same time, Proposition~\ref{prop:debunch} implies that the probability of debunching is
\begin{equation}\label{eq:debunch_FZ}
\Prb(R\leq1+\tau)
=
F_Z\!\left(\log(1+\tau)\right).
\end{equation}
Equations~\eqref{eq:stoploss} and~\eqref{eq:debunch_FZ} show that debunching and average log threshold response impose joint restrictions on the same valuation-overshoot distribution. A distribution of overshoots that explains how much integer mass disappears must also explain the magnitude of the average log threshold response. Observing more than one tax rate therefore places multiple restrictions on the same latent distribution. Section~5 studies what these restrictions identify when only a finite set of tax rates is observed.

\section{Identification}\label{sec:identification}

Section~\ref{seccrosssectionimplications} established that the no-tax integer discontinuity
\[
J_n(0)=Q_0(n-)-Q_0(n)
\]
identifies the mass of consumers with a binding threshold at $n$, $\Prb(T=n,V>n)=J_n(0)$. Ordinary no-tax demand does not identify the distribution of $V$ above a binding threshold. We now ask what tax variation adds.

\subsection{Identification from individual cutoff data}

The within-subjects experimental design contains substantially more information about binding threshold types. For a consumer with no-tax cutoff $C$, observing how that cutoff changes under a known tax can reveal information about that consumer's willingness to pay.

\begin{proposition}[Point identification for threshold movers]\label{prop:paired}
Fix a tax rate $\tau>0$. Suppose we observe a consumer's pre-tax posted price cutoff without tax, $C_A(0)=C$, and with tax, $C_A(\tau)$. The following statements hold almost surely:
\begin{enumerate}[label=(\roman*)]
\item If $C$ is noninteger, then $V=C$.
\item If $C=n\in\N_+$ and $C_A(\tau)<n$, then the consumer is a binding threshold-$n$ mover and satisfies $T=n$ and $V=(1+\tau)C_A(\tau)$.
\item If $C=n\in\N_+$ and $C_A(\tau)=n$, then the consumer stays at the binding threshold with $T=n$ and $V\ge n(1+\tau)$.
\end{enumerate}
\end{proposition}

Part $(ii)$ is the key identification gain from tax variation. An integer no-tax cutoff reveals only $V>n$; observing that cutoff move under a known tax reveals the no-tax willingness to pay of a mover. Part $(iii)$ shows that we can attain a tighter lower bound for the willingness to pay of consumers who do not change their cutoff in response to a tax. For a binding threshold-$n$ consumer, the no-tax cutoff implies only $V>n$, whereas remaining at $n$ under tax $\tau$ implies $V\ge n(1+\tau)$.

\subsection{Identification from population moments}

Proposition~\ref{prop:paired} uses individual cutoff changes to learn about individual willingness to pay. We now consider what can be learned about the distribution of the valuation overshoot for binding-threshold consumers from the two population moments derived in Section~\ref{sec:overshoot_behavior}. At a given tax rate, the debunching probability identifies a point on the CDF of $Z=\log R$, while the mean log threshold response identifies the area under the CDF up to that point. Observing these moments at multiple tax rates therefore places multiple restrictions on the same overshoot distribution.

\begin{proposition}[Identification from log threshold response and debunching]\label{prop:partialover}
Let $\Lambda$ denote the set of positive tax rates at which these population moments are available. For any binding-threshold population and each $\tau\in\Lambda$, the debunching probability identifies the share of consumers with valuation overshoot no greater than $1+\tau$, i.e.,
\[
1-\frac{\sum_{n\in\mathbb N_+}J_n(\tau)}
        {\sum_{n\in\mathbb N_+}J_n(0)}
=F_Z\!\left(\log(1+\tau)\right),
\]
while the mean log threshold response identifies the average log distance of debunching consumers' overshoots below $1+\tau$, weighted by the share who debunch, i.e.,
\[
\tau\,\E[\phi_A(\tau)]=\int_0^{\log(1+\tau)}F_Z(z)\,\dd z.
\]
The identified set for $F_Z$ consists of all distributions satisfying the model assumptions and these moment restrictions for every $\tau\in\Lambda$.  
\end{proposition}

With richer tax variation, these restrictions become more informative. If the mean log threshold response were observed over a continuum of tax rates, differentiating its integral representation in equation~\eqref{eq:stoploss} would recover the CDF over the corresponding range. Even more directly, observing the debunching probabilities over a continuum of tax rates would trace out the overshoot distribution. With only finitely many moment restrictions, many overshoot distributions can satisfy the same CDF values and areas under the CDF, which is why the distribution is only partially identified from these moments. 

The identification result in Proposition~\ref{prop:partialover} is limited in one more way. Even with a continuum of tax rates, the binding-threshold population moments identify the distribution of log $R$, and equivalently $R$. They generally do not identify the joint distribution of the model primitives $T$ and $V$. This poses a challenge for welfare analysis, as we discuss in Section~\ref{sec:welfare}.

\section{Estimation strategy}\label{sec:estimation}

The experiment provides reported posted-price cutoffs elicited by the BDM mechanism. Our empirical strategy connects these reports to the model.

\subsection{Estimation sample}

We use the publicly available replication data from \citet{TRJ2018}. Recall that $P^1_{ik}$ denotes consumer $i$'s Module 1 reported posted-price cutoff for product $k$, while $P^2_{ik}$ denotes the corresponding untaxed Module 2 cutoff. Throughout the empirical analysis, an observation refers to a consumer-product pair, for which we observe the two reported cutoffs $P^1_{ik}$ and $P^2_{ik}$. In the standard-tax and triple-tax arms, $P^1_{ik}$ is elicited under the randomized tax treatment; in the no-tax arm, both cutoffs are untaxed. These cutoffs equal the incentivized reports unless a report is censored at the upper endpoint, in which case we incorporate the consumer's follow-up response for the maximum tag price she is willing to pay.\footnote{\citet{TRJ2018} elicited the incentivized reports using sliders with a finite upper endpoint. When a report reached that endpoint, respondents answered a follow-up question about the maximum tag price they were willing to pay. The replication variables used here incorporate that follow-up response.}

We restrict the sample to consumers who correctly answer the primary Module 1 tax-comprehension questions. This reduces the sample from 4,223 to 2,998 consumers. Following \citet{TRJ2018}, we additionally restrict attention to consumer-product observations satisfying $P^2_{ik}\ge1$, leaving 58,478 observations from 2,991 consumers. Analyses restricted to the untaxed Module 2 distribution in the no-tax arm use 21,551 observations from 1,098 consumers. Since each consumer reports cutoffs for twenty products, bootstrap inference treats the consumer rather than the consumer-product observation as the sampling unit. Appendix~\ref{app:sample_cells} gives the complete sample construction.

\subsection{Integer reports correspond to threshold mass}\label{sec:direct_source}

Section~\ref{seccrosssectionimplications} showed that a positive-integer discontinuity in no-tax demand is generated by binding threshold types. To connect this implication to the reported BDM cutoffs, we focus on exact integer reports in untaxed Module 2 over
\[
\mathcal N_I:=\{2,3,\ldots,14\}.
\]
Integer 1 is excluded because the preferred sample is truncated at $P^2\ge1$, while we stop at 14 to avoid the upper response boundary at 15.

\begin{assumption}[True integer reports]\label{ass:directinteger}
For each $n\in\mathcal N_I$, an untaxed Module 2 report satisfies $P^2_{ik}=n$ if and only if consumer $i$, for product $k$, belongs to the population mass generating the no-tax demand discontinuity
\[
J_n(0)=Q_0(n^-)-Q_0(n).
\]
\end{assumption}

Under true integer reports, the share of reports at an integer corresponds to the mass generating the no-tax discontinuity at that integer,
\[
\Prb(P^2_{ik}=n)=J_n(0).
\]
Proposition~\ref{propnotaxid} shows that this discontinuity identifies the binding threshold-$n$ mass,
\[
J_n(0)=\Prb(T=n,V>n).
\]
Thus, the mass of Module 2 integer reports is the model's binding-threshold mass. We estimate the binding-threshold share separately in each arm.

\subsection{Tax response}\label{sec:response_moments}

Following \citet{TRJ2018}, define the log change in the reported cutoff for consumer $i$ and product $k$ by
\begin{equation}\label{eq:y_response}
y_{ik}:=\log P^2_{ik}-\log P^1_{ik}.
\end{equation}
We use log changes rather than proportional changes for the empirical response moment because, when reported cutoffs are small, modest reporting noise can generate very large values of $P^2_{ik}/P^1_{ik}$. Taking logs compresses such ratios and limits their influence on the moment.

A challenge for estimating the tax response is that there is substantial variation in reports between the two modules even in the no-tax arm, with Module 2 reports being 37 cents lower ($p<0.001$) on average than Module 1 reports (\textit{order effects}). Thus $y_{ik}$ in the tax arm contains both the response to the tax and any order effects from eliciting Module 2 after Module 1. The following assumption will allow us to estimate the order effects from the no-tax arm and use them to isolate the cutoff change arising from taxes in the tax arms. 

\begin{assumption}[Stable order effects]\label{ass:ordereffects}
Mean order effects do not vary across tax rates (thus across experimental arms). This assumption applies separately to the binding-threshold population and the remaining population.
\end{assumption}

Let $\xi_A$ and $\xi_C$ denote the mean no-tax log cutoff changes for the binding-threshold consumers and continuous residual consumers, respectively. Let $a_i\in\{0,S,H\}$ denote consumer $i$'s experimental arm, where $0$ denotes no tax, $S$ standard tax, and $H$ triple tax. Let $m_{A,a}$ and $m_{C,a}$ denote their mean order-adjusted log tax responses in treated arm $a$. We estimate each in the following way: subtract the group's no-tax mean log change from each treated observation's log change, divide by the experimental tax rate $\tau_{i}$, and average within that group. Thus,
\begin{equation}\label{eq:response_decomp_main}
\begin{aligned}
m_{A,a}&:=\E\!\left[\frac{y_{ik}-\xi_A}{\tau_i}\,\middle|\,P^2_{ik}\in\mathcal N_I,a_i=a\right],\\
m_{C,a}&:=\E\!\left[\frac{y_{ik}-\xi_C}{\tau_i}\,\middle|\,P^2_{ik}\notin\mathcal N_I,a_i=a\right].
\end{aligned}
\end{equation}
When tax rates are small, $m_{A,a}$ and $m_{C,a}$ approximate the average behavioral tax responses, $\psi_A$ and $\theta$, in the binding-threshold and continuous residual populations, respectively.

\subsection{Threshold persistence and gross debunching}\label{sec:three_deheap}

Section~\ref{sec:hybrid_tax} showed that a binding threshold-$n$ consumer debunches when her taxed cutoff falls below $n$. Our empirical target is thus the fraction of the original binding threshold-$n$ population whose cutoff leaves $n$ because of the tax. We refer to this as \emph{gross debunching} because it counts all movement away from a threshold rather than the net change in mass at that threshold, which would include those whose cutoff becomes $n$ after tax.

For treated arm $a\in\{S,H\}$, define the gross debunching probability at threshold $n$ by
\[
d^{\mathrm{gross}}_{a,n}
:=\Prb\!\left(V<n(1+\tau_i)\mid T=n,V>n,a_i=a\right).
\]
This is the probability that a binding threshold-$n$ type's model cutoff leaves $n$ under the tax.

Reported cutoffs create an additional complication. Assumption~\ref{ass:directinteger} classifies binding threshold types using the untaxed Module-2 report, but the corresponding Module 1 report need not equal the same integer even in the absence of an experimental tax (recall our order effects discussion). Let $q_{0,n}$ denote the no-tax \emph{same-$n$ persistence rate}: the probability that a no-tax observation with Module 2 report $n$ is again reported at $n$ in Module 1. Let $q_{a,n}$ denote the corresponding same-$n$ persistence rate in treated arm $a$. Even in the no-tax arm, same-$n$ persistence is well below one. 

We use the no-tax persistence rate as a benchmark for order effects when estimating how much of the treated-arm decline in persistence reflects structural debunching. Turning the observed persistence rates into the structural gross debunching probability requires one additional reporting assumption.

\begin{assumption}[Threshold report persistence]\label{ass:source_measurement} For each treated arm $a\in\{S,H\}$ and threshold $n\in\mathcal N_I$:
\begin{enumerate}
    \item Binding threshold-$n$ consumers who debunch under the model report $n$ in Module 1 with probability zero.\footnote{If some structural debunchers nevertheless report $n$ in the tax arm, then this will lead to underestimation of true gross debunching.} 
    \item Binding threshold-$n$ consumers who do not debunch report $n$ in Module 1 with the same probability as binding threshold-$n$ consumers in the no-tax arm.
\end{enumerate}
\end{assumption}
Assumption~\ref{ass:source_measurement} implies the following relationship: for every treated arm $a\in\{S,H\}$ and $n\in\mathcal N_I$ with $q_{0,n}>0$,
\begin{equation}\label{eq:persistence_factorization}
q_{a,n}=q_{0,n}\bigl(1-d^{\mathrm{gross}}_{a,n}\bigr).
\end{equation}
Rearranging, our estimate for gross debunching in tax arm $a$ at integer $n$ is 
\begin{equation}\label{eq:grossdeheapn}
\widehat d^{\mathrm{gross}}_{a,n}
=1-\frac{\widehat q_{a,n}}{\widehat q_{0,n}}.
\end{equation}

To aggregate across thresholds, let $w_{a,n}$ denote the fraction of treated arm $a$'s binding-threshold observations whose Module 2 report is $n$. Our aggregate estimate is
\begin{equation}\label{eq:grossdeheapagg}
\widehat d^{\mathrm{gross}}_a
=\sum_{n\in\mathcal N_I}\widehat w_{a,n}\widehat d^{\mathrm{gross}}_{a,n}.
\end{equation}
Thus, each threshold-specific estimate is weighted by that threshold's share of the treated arm's binding-threshold population.

For comparison, we report also the \emph{net integer-mass loss}. Within each arm, we subtract the Module-1 integer share from the Module-2 integer share and divide by the Module-2 integer share. We then subtract the resulting no-tax rate from the treated rate. This comparison assumes that, absent the tax effect, the proportional change in integer mass would be the same across arms. Net integer-mass loss is distinct from our structural debunching probability because a decline in mass at one integer can be offset by movement into another integer. Hence, we prefer to use gross debunching for our overshoot analysis in the next section. We report also the decline in same-$n$ persistence: the no-tax fraction of binding-threshold observations with identical reports across modules minus the corresponding treated-arm fraction. Appendix~\ref{app:source_deheap} contains the details for this subsection.

\subsection{Overshoot distribution}\label{sec:overshoot_gmm}

Proposition~\ref{prop:partialover} provides log overshoot identification results from debunching and log threshold-response moments. Since the logarithm is a one-to-one transformation, the identification result carries through to $R=V/T$. We work directly with $R$ and obtain identified sets for its CDF $F_R$. 

The experiment contains many local sales tax rates. However, we have too few observations per tax rate to do such granular analysis. We instead use two aggregate moments in each tax arm. The four empirical targets are the mean log threshold response and gross debunching probability in each treated arm: $\widehat m_{A,S}$, $\widehat d^{\mathrm{gross}}_S$, $\widehat m_{A,H}$, and $\widehat d^{\mathrm{gross}}_H$. The central empirical restriction is that the same overshoot distribution must jointly explain all four moments.

For a candidate overshoot $r$, the two structural functions are those derived in Section~\ref{sec:hybrid_tax}:
\begin{equation}\label{eq:overshoot_functions}
\phi_A(\tau_i,r)=\frac{[\log(1+\tau_i)-\log r]_+}{\tau_i},
\qquad
\1\{r<1+\tau_i\}.
\end{equation}
The first gives the log threshold response generated by a type with overshoot $r$, while the second indicates whether that type debunches. We impose a common overshoot distribution across the experimental arms and local sales-tax rates. For each candidate overshoot distribution, we calculate the predicted mean log threshold response and debunching probability at each observed tax rate. We then average these predictions within each treated arm, weighting each tax rate by its share of that arm’s binding-threshold observations.

We first ask whether any candidate distribution can reproduce all four estimated moments exactly. Conditional on feasibility, we obtain bounds for $F_R(r)$ by minimizing and maximizing the fraction with overshoot no greater than $r$. The moments and CDF values are linear in the probabilities assigned to candidate overshoots. Appendix~\ref{app:commonovershoot} gives the details.

\section{Empirical results}\label{sec:results}

First, we provide an estimate of the binding-threshold mass. Next, we estimate tax response and gross debunching for those consumers. Then, we test restrictions on how consumers' reports change with taxes. Finally, we ask whether tax response and gross debunching can be jointly rationalized by an overshoot distribution and provide identified sets for the distribution.

\subsection{Binding-threshold mass}

Under Assumption~\ref{ass:directinteger}, the fraction of untaxed Module 2 integer reports corresponds directly to the mass of binding-threshold consumers. Recall that we restrict our attention to integers 2-14, and each observation is a consumer-product report. In the no-tax arm, this amounts to 2,803 of 21,551 total observations, giving a binding-threshold mass of 13 percent. The consumer-cluster bootstrap standard error is 0.72 percentage points and the $95\%$ confidence interval is $[11.62\%,14.44\%]$. 

\begin{figure}[H]
\centering
\includegraphics[width=0.88\textwidth]{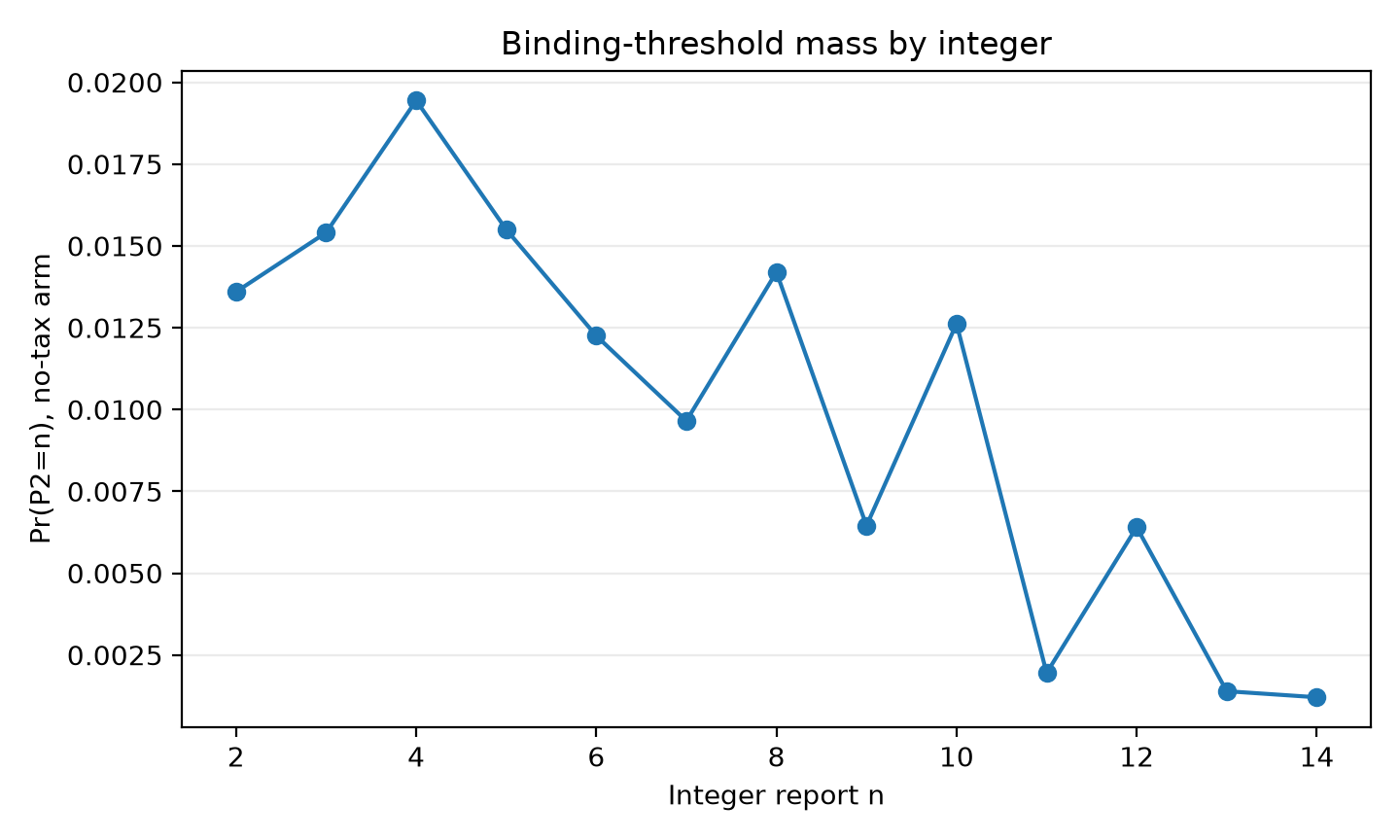}
\caption{Binding-threshold mass by integer in the no-tax arm. The figure plots the share of untaxed Module-2 reports equal to each integer $n\in\{2,\ldots,14\}$. Under Assumption~\ref{ass:directinteger}, each point measures the binding threshold-$n$ mass $\Pr(T=n,V>n)$. The masses sum to 0.1301.}
\label{fig:byinteger}
\end{figure}

Figure~\ref{fig:byinteger} shows the mass integer by integer. There is substantial variation in mass across thresholds, ranging from $0.0012$ at $\$14$ to $0.0194$ at $\$4$. Given the small number of observations at each integer price, we will focus on arm-wide moments for our identification results. 

Our integer support restriction on attention thresholds was motivated by evidence on demand discontinuities at integer prices. We check whether there are masses at other focal prices as well. In the no-tax Module 2, the share of reports at half-dollar prices over $2.50,3.50,\ldots,13.50$ is $0.0206$, or $15.8 $ percent of the integer share. Thus, we retain our focus on integers thresholds.

We estimate the binding-threshold share separately in each arm. Table~\ref{tab:armstable} reports $13.01\%$ in the no-tax arm, $13.62\%$ in the standard arm, and $11.96\%$ in the triple arm. Relative to the no-tax arm, the differences are 0.61 percentage points for standard arm ($p=0.571$) and -1.04 percentage points for triple arm ($p=0.322$). Neither difference is statistically distinguishable from zero. 

\begin{table}[H]
\centering
\caption{Share of integer reports in Module 2 across arms}
\label{tab:armstable}
\footnotesize
\setlength{\tabcolsep}{4pt}
\begin{tabular}{lrrr}
\toprule
Arm & Share of integer reports& Difference from no tax & Two-sided $p$-value \\
\midrule
No tax   & 0.1301 & --- & --- \\
Standard & 0.1362 & +0.0061 & 0.571 \\
Triple   & 0.1196 & -0.0104 & 0.322 \\
\bottomrule
\end{tabular}
\begin{minipage}{0.93\textwidth}\footnotesize
Notes: Each row gives the share of Module-2 reports that are integers from 2 through 14. Differences are relative to the no-tax arm; neither the standard-arm difference nor the triple-arm difference is statistically distinguishable from zero. The $p$-values use a normal approximation with consumer-cluster bootstrap standard errors based on 5,000 draws. Appendix Figure~\ref{fig:arm} plots these estimates along with 95-percent confidence intervals.
\end{minipage}
\end{table}

\subsection{Tax response and gross debunching}

Table~\ref{tab:gmmest} provides estimates for the log tax response, separately for binding-threshold consumers and the remaining continuous-residual consumers. Average threshold response is 0.269 in the standard tax arm and 0.383 in the triple tax arm. Continuous-residual response is 0.247 under the standard tax and 0.488 under the triple tax, close to the original \citet{TRJ2018} estimates of 0.250 and 0.475. For small tax rates, the tax response can be interpreted as an estimate of average $\psi_A$ and $\theta$, as discussed in Section~\ref{sec:response_moments}. Intuitively, a consumer with a tax response of 0.25 has a quarter of the response of an optimizing consumer. These estimates are consistent with the model's prediction that consumers under-react to taxes.

\begin{table}[H]
\centering
\caption{Tax response by group}
\label{tab:gmmest}
\small
\begin{tabular}{lccc}
\toprule
Arm & Binding share & Continuous-residual response& Threshold response\\
\midrule
Standard & 0.1362 & 0.2466$^{**}$ (0.1015) & 0.2685 (0.1798) \\
Triple   & 0.1196 & 0.4876$^{***}$ (0.0417) & 0.3829$^{***}$ (0.0719) \\
\bottomrule
\end{tabular}
\begin{minipage}{0.94\textwidth}\footnotesize
Notes: Tax response is measured as the order-adjusted log change in the reported cutoff divided by the observation's experimental tax rate. The binding share is the fraction of the treated arm's Module 2 reports at integers from 2 through 14. The threshold response point estimates are directionally consistent with increasing behavioral response, but the difference is not statistically significant. Bootstrap standard errors from 5,000 consumer-cluster draws are in parentheses. $^{*}p<0.10$, $^{**}p<0.05$, and $^{***}p<0.01$ for two-sided normal-approximation tests against zero using the bootstrap standard errors.
\end{minipage}
\end{table}
The threshold response estimates are imprecise, likely due to the smaller sample of binding-threshold consumers. We therefore cannot conclude that their response is statistically significantly higher in the triple-tax arm. Nevertheless, the point estimates rise with the tax rate, matching the pattern in the continuous residual. Moreover, our empirical specification measures the log response, which understates the behavioral response. For every binding threshold type, the log tax response is weaker than the behavioral tax response, $\phi_A\leq\psi_A$, with strict inequality when the cutoff moves. For a fixed type, the gap increases with the tax rate. Thus, the estimated increase occurs despite the greater compression of behavioral responses at larger taxes. Our estimates are directionally consistent with stronger behavioral responses to larger taxes, although the comparison remains statistically inconclusive.

Table~\ref{tab:three_deheap} presents our estimates of how binding-threshold shares change in response to taxes. The net integer-mass loss is 31.5 percent in the standard arm and 25.1 percent in the triple arm. Recall these are net changes and need not equal the fraction of original threshold types that leave their own integer (since some may move their report to another integer threshold). Thus, net integer-mass loss is a reduced-form measure of debunching, and our estimates are consistent with the model prediction of mass loss at integer prices.

Next, we study what share of consumers change their cutoff in response to the tax. Even in the no-tax arm, binding-threshold observations have different reports between Modules 1 and 2. Therefore, we measure same-\textit{n} persistence relative to the no-tax arm. Same-$n$ persistence falls from 0.328 in the no-tax arm to 0.190 in the standard arm and 0.155 in the triple arm. Under Assumption~\ref{ass:source_measurement}, our estimates imply that about 41.5 percent of the binding-threshold population debunches in the standard arm and 52.2 percent in the triple arm.

\begin{table}[H]
\centering
\caption{Net integer-mass loss, threshold persistence, and gross debunching}
\label{tab:three_deheap}
\footnotesize
\setlength{\tabcolsep}{4pt}
\begin{tabular}{lccc}
\toprule
Arm & Net integer-mass loss & Same-$n$ persistence decline & Gross debunching \\
\midrule
Standard & 0.3148$^{***}$ (0.0499) & 0.1376$^{***}$ (0.0213) & 0.4148$^{***}$ (0.0569) \\
Triple   & 0.2511$^{***}$ (0.0525) & 0.1723$^{***}$ (0.0204) & 0.5220$^{***}$ (0.0522) \\
\bottomrule
\end{tabular}
\begin{minipage}{0.95\textwidth}\footnotesize
Notes: Net integer-mass loss is the treated arm's proportional decline from Module 2 to Module 1, minus the corresponding no-tax proportional decline. The same-$n$ persistence decline compares how often an observation reported at threshold $n$ in Module 2 is reported again at $n$ in Module 1. Gross debunching additionally uses Assumption~\ref{ass:source_measurement} to provide our tax-induced model debunching estimate. Bootstrap standard errors are in parentheses, using 5,000 consumer-cluster draws (4,995 valid draws for gross debunching). $^{*}p<0.10$, $^{**}p<0.05$, and $^{***}p<0.01$ for two-sided normal-approximation tests against zero using the bootstrap standard errors.
\end{minipage}
\end{table}

We find that integer masses shrink with taxes. The share of Module 1 cutoffs at the same integer as Module 2 cutoffs in the tax arms declines even after adjusting for order effects using the no-tax arm. Finally, we find that roughly half the binding-threshold population debunches under taxation. Notice the net integer-mass loss estimate is lower for the triple tax arm compared to the standard tax arm (although statistically indistinguishable with $p=0.267$), while the gross debunching estimate is higher ($p=0.125$). As we show in the next section, there is significant integer to integer movement in reports in the triple tax arm, which is contributing to this phenomenon. Although our standard errors are large again, we find that the gross debunching estimates fit the pattern of greater debunching with higher taxes. Thus, the debunching estimates are consistent with the attention threshold mechanism. 

\subsection{Atom relocation and mover support}\label{sec:nonewatoms_empirical}

Proposition~\ref{prop:nonewatoms} predicts that mass leaving a binding threshold should disperse below the threshold in a certain range rather than reappearing as a new point mass. Since reported cutoffs are observed up to the cent, we test for a tax-induced excess mass at a noninteger cent price. For each noninteger price between \$1.25 and \$14.75, we first measure excess mass at that price relative to neighboring noninteger cent bins, using a local quadratic to construct the neighboring-price benchmark. We then compare the tax module local excess mass with the corresponding no-tax local excess mass, expressing each as a share of that arm's binding-threshold observations. Appendix~\ref{app:noninteger_scan} gives the details.

The scan covers 1,338 noninteger cent prices and controls the familywise error rate within each treated arm. In the standard arm, the largest positive excursion is $z=2.445$ at \$10.50 (adjusted $p=0.994$). In the triple arm, the largest excursion is $z=3.346$ at \$3.99 (adjusted $p=0.262$). Thus, we find no evidence of a tax-induced spike at noninteger cutoffs, which is consistent with the attention threshold model.

The model also rules out tax-induced relocation into a different integer (since $V$ is atomless and attention thresholds are static). Relative to the no-tax arm, the excess probability of landing at any lower integer in the standard arm is $-0.83$ percentage points ($p=0.749$), while excess $n\to n-1$ movement is $-0.14$ percentage points ($p=0.546$). In the triple-tax arm, the excess probability of landing at a lower integer is 5.41 percentage points ($p=0.0004$), including 2.96 percentage points of excess $n\to n-1$ movement ($p=0.0018$). These are one-sided tests of positive excess relocation. Our estimates of movement of cutoffs to other integers in response to the consumers' local sales taxes (standard tax arm) are consistent with the model's predictions; we do not find significant excess movement to lower integers. However, we do see some integer mass relocate to another integer in response to tripling the local sales tax. This pattern is consistent with the triple tax arm having lower estimated net integer-mass loss but higher estimated gross debunching than the standard arm. The model cannot explain this relocation to another integer as a change in true cutoffs.

A separate paired diagnostic uses the model's mover support restriction $C_A(\tau_i)>n/(1+\tau_i)$. The posted-price cutoff when there is a tax must satisfy a lower bound implied by the tax rate and no-tax threshold cutoff. Among binding-threshold observations, 22.61 percent of standard-arm reports and 24.05 percent of triple-arm reports cross this bound. However, we are hesitant to interpret this as a violation of the model.

Evaluating no-tax reports at the corresponding hypothetical tax bounds gives 16.91 and 10.06 percent, respectively. Thus, there are violations of the lower bound even without taxation. Among binding-threshold observations in the no-tax arm, removing the average log order effect yields crossing rates of 29.40 percent and 13.38 percent at the hypothetical standard- and triple-tax lower bounds, respectively. These crossings occur despite neither module imposing a tax. A model-consistent tax response moves cutoffs toward the lower bound, so reporting noise can generate additional crossings in the tax arms. Consequently, this data does not allow us to cleanly test the model's lower bound on movers' taxed cutoffs.

\subsection{Joint fit and overshoot identification}\label{sec:overshoot_pi}

In this section, we ask whether the same overshoot distribution of $R=V/T$ can jointly explain log threshold response and gross debunching moments in both tax arms. The common $R$ specification reproduces all four estimated moments. This allows us to partially identify the overshoot distribution. 

Table~\ref{tab:overshoot} reports identified sets for the binding-type CDF $F_R$ under the common overshoot specification.

\begin{table}[H]
\centering
\caption{Identified sets for the binding-type overshoot CDF}
\label{tab:overshoot}
\small
\begin{tabular}{rc}
\toprule
$r$ & Common $F_R$ \\
\midrule
1.03 & [0.101, 0.411] \\
1.05 & [0.222, 0.411] \\
1.07 & [0.321, 0.493] \\
1.10 & [0.438, 0.522] \\
1.15 & [0.496, 0.522] \\
1.21 & [0.503, 0.533] \\
1.25 & [0.506, 0.593] \\
\bottomrule
\end{tabular}
\begin{minipage}{0.94\textwidth}\footnotesize
Notes: $F_R(r)$ is the share of binding-threshold types whose valuation overshoot satisfies $V/T\le r$. The common specification imposes one overshoot distribution across experimental arms and local sales-tax rates. The intervals give the minimum and maximum CDF values across distributions satisfying the four estimated moments. Appendix~\ref{app:commonovershoot} gives the construction.
\end{minipage}
\end{table}

At $r=1.10$, the common specification gives $F_R(1.10)\in[0.438,0.522]$. Thus, between 43.8 and 52.2 percent of binding threshold types have willingness to pay no more than 10 percent above their threshold. Recall that we do not restrict the dependence between $V$ and $T$ and imposing such restrictions can sharpen our identification result. Nevertheless, the restrictiveness of the attention mechanism, in addition to providing sharp predictions for behavior, enables meaningful identification of willingness to pay relative to the attention threshold. Bunching at an integer cutoff pins down the attention threshold. At that threshold and given a tax rate, heterogeneity in tax responses and debunching must arise from heterogeneity in $V$. Since we look at aggregate binding-threshold data (thus, don't use tax response and debunching moments at each threshold separately), we cannot identify the joint distribution of $V$ and $T$, but we can partially identify the distribution of their ratio. 

\section{Welfare Analysis}\label{sec:welfare}

The same no-tax and taxed cutoffs can have different welfare implications depending on whether the no-tax cutoff arises from preference or attention. We compare the RUM--APT and TRJ welfare interpretations for a binding-threshold consumer whose no-tax and taxed cutoffs can be rationalized by both models. Then, we use the partially identified overshoot distribution from Section~\ref{sec:overshoot_pi} to estimate bounds on the welfare loss from taxes.

Fix an ad valorem tax rate $\tau>0$ and producer price $p\ge0$. As in \citet{TRJ2018}, tax revenue is treated as an efficient transfer and $p$ is marginal social cost under perfectly elastic supply. We define the \emph{excess burden of taxation} as the reduction in total surplus caused by introducing the tax relative to the consumer's no-tax choice. A purchase generates social surplus equal to willingness to pay minus $p$, while not purchasing generates zero surplus. Thus, when the tax causes a consumer whose willingness to pay exceeds $p$ to stop purchasing a good, excess burden equals the lost surplus from that transaction.

\subsection{RUM--APT and TRJ can rationalize the same binding-threshold behavior}

Fix a tax rate $\tau>0$ and consider a binding-threshold consumer. Her no-tax cutoff is $C_A(0)=C=T$ and $V>T$. If the consumer moves under tax $\tau$, then $C_A(\tau)<C_A(0)$ and Proposition~\ref{prop:paired} gives
\[
V=(1+\tau)C_A(\tau)>C_A(0).
\]
If the consumer stays at the same threshold, then $V\ge(1+\tau)C$. By Proposition~\ref{prop:thresholdresponse}, we know that the tax response, $\psi_A(\tau)$, is bounded with $0\leq\psi_A(\tau)<1$. 

The same no-tax and taxed cutoffs can also be rationalized by the TRJ model. Under TRJ, the no-tax cutoff is the willingness to pay, so set
\begin{equation}\label{eq:trj_wtp}
V^{\mathrm{TRJ}}:=C_A(0).
\end{equation}
Next, let the TRJ tax response be
\begin{equation}\label{eq:trj_theta}
\theta^*:=\frac{C_A(0)/C_A(\tau)-1}{\tau}=\psi_A(\tau).
\end{equation}
Then, we get that
\[
C_{\mathrm{TRJ}}(0)=V^{\mathrm{TRJ}}=C_A(0)
\qquad\text{and}\qquad
C_{\mathrm{TRJ}}(\tau)=\frac{V^{\mathrm{TRJ}}}{1+\theta^*\tau}=C_A(\tau).
\]
Thus, if the cutoff behavior of a binding-threshold individual can be rationalized by RUM-APT, then it can be rationalized by TRJ as well. 
%\begin{proposition}[TRJ representation of binding-threshold behavior]\label{prop:trjrepresentation}
%Fix a tax rate $\tau>0$. Every RUM-APT binding-threshold consumer has tax response
%\[
%0\le\theta^*<1.
%\]
%Hence the same no-tax and taxed cutoffs can be generated by the TRJ model with willingness to pay $V^{\mathrm{TRJ}}=C_A(0)$ and tax response $\theta^*$.
%\end{proposition}

The TRJ model is more flexible in two ways. First, the TRJ model can explain overreaction to taxes with $\theta>1$, while binding-threshold consumers under the RUM--APT model necessarily underreact to taxes. Second, the RUM-APT model places restrictions on the tax response of binding-threshold consumers governed by equation~\eqref{eq:psiA_formula} while in the TRJ model, the tax response is defined behaviorally by $\theta\ge0$, which can vary with the tax rate. 

However, at the population level there is an important difference. The TRJ model assumes that demand is smooth, so it cannot explain demand discontinuities while the RUM--APT model rationalizes them using attention thresholds. We do not explore the full empirical content of both models, so these differences need not be exhaustive. When comparing to the TRJ model, we will adopt the TRJ interpretation that binding-threshold consumers have willingness to pay equal to their no-tax integer cutoffs, even if the resulting population does not satisfy their smooth demand assumption.\footnote{Figure 3 in \citet{TRJ2018} plots average demand curves across all twenty products, with price on the
vertical axis and quantity on the horizontal axis.
Masses of reported cutoffs at integer prices appear
as horizontal line segments at those prices.} 

\subsection{Excess burden of taxation under both models}

The tax eliminates a purchase when
\[
C_A(\tau)\le p<C_A(0).
\]
Under the TRJ interpretation, $V^{\mathrm{TRJ}}=C_A(0)$, so individual excess burden for consumers who stop choosing the good due to the tax is
\begin{equation}\label{eq:eb_trj}
\mathrm{EB}_{\mathrm{TRJ}}(p)
:=C_A(0)-p.
\end{equation}
Under RUM--APT, individual excess burden for those consumers is
\begin{equation}\label{eq:eb_apt}
\mathrm{EB}_{\mathrm{APT}}(p)
:=V-p.
\end{equation}
For a binding-threshold consumer who stops choosing the good, $C_A(0)=T$ and $V>C_A(0)$, so
\begin{equation}\label{eq:welfare_gap}
\mathrm{EB}_{\mathrm{APT}}(p)-\mathrm{EB}_{\mathrm{TRJ}}(p)
=V-T>0.
\end{equation}
For a consumer whose choice is unaffected by the tax, the excess burden under both models is zero. These consumers can be divided into two groups: always buyers and never buyers. Always buyers buy with and without the tax. They lose surplus equal to the tax, but the tax is a pure transfer, so it does not contribute to excess burden of taxation. Consumers who never buy, with or without tax, lose nothing due to the tax. For those who do change their choice, Proposition ~\ref{prop:paired} identifies $V=(1+\tau)C_A(\tau)$. Thus, the excess burden of taxation for binding-threshold consumers is identified from individual-level cutoff data in the RUM-APT model, even though we do not identify the willingness to pay of every such consumer. 

Let $G_{\mathrm{APT}}$ and $G_{\mathrm{TRJ}}$ denote the CDFs of excess burden across the population of binding-threshold consumers under the RUM--APT and TRJ interpretations, respectively.

\begin{proposition}[Welfare comparison]\label{prop:welfaredom}
For every $z\in\mathbb R$,
\begin{equation}\label{eq:fosd}
G_{\mathrm{APT}}(z)\le G_{\mathrm{TRJ}}(z).
\end{equation}
Thus, the distribution of excess burden under RUM--APT first-order stochastically dominates the distribution of excess burden under TRJ. If a positive mass of binding-threshold consumers have their purchase eliminated by the tax at producer price $p$, the dominance is strict for some $z$.
\end{proposition}

The two models agree on the excess burden from consumers who do not change their choice. They disagree on the value of the lost surplus for those who stop choosing the good. TRJ interprets the no-tax cutoff as willingness to pay, $V^{\mathrm{TRJ}}=C_A(0)=T$. RUM--APT instead interprets the same cutoff as an attention threshold, and for a consumer whose purchase is eliminated, the taxed cutoff reveals $V>T$. The same behavioral response therefore corresponds to a larger loss of surplus under RUM--APT. This parallels the welfare analysis logic in \citet{Patel2025} for the equivalent variation of a price change (see Theorem 5.3 in that paper). 

In this paper, our comparison concerns the welfare consequences of taxation. Since \citet{Ramsey1927}, it has been understood that demand elasticity plays an important role in the welfare effects of commodity taxation. The conclusion arising from \citet{CLK2009} and \citet{TRJ2018} that taxes may have lower welfare costs than implied by the neoclassical model rests on the claim that demand response to taxes is weaker than under full optimization.\footnote{\citet{TRJ2018} also show that when there is heterogeneity in attention to taxes, welfare costs may be higher than in the homogeneous attention model of \citet{CLK2009}.} However, the RUM--APT assigns a greater welfare loss to the consumers who stop purchasing; thus, a smaller tax response need not imply a smaller welfare loss. 

In fact, even with a weaker demand response to taxes than implied by the neoclassical model, the welfare losses may be larger. Suppose there is a good with a price of $\$4.95$, and there are two consumers with no-tax cutoffs of $\$5$. Both consumers choose the good. The neoclassical model would interpret the five dollar cutoff as willingness to pay without taxes, while for binding-threshold consumers the willingness to pay is greater than five dollars. Suppose a tax of $10\%$ is introduced. With the tax, both consumers under the neoclassical model would stop choosing the good, and the total surplus lost would be ten cents. Suppose instead there is a weaker response and only one consumer changes her choice. Under RUM--APT, we can rationalize this behavior with one consumer having a willingness to pay of $\$5.40$ and another of $\$6$. The resulting lost surplus---and excess burden---is $45$ cents and exceeds that implied by the neoclassical model despite fewer consumers changing their choice.  

Perfectly elastic supply keeps the producer price fixed when the tax is introduced. If the producer price were to fall, welfare analysis under RUM–APT would also need to account for consumers who begin considering the good because its posted price falls below their attention thresholds.

\subsection{Bounds on excess burden}\label{sec:welfare_relative}

When a tax is introduced, a binding-threshold consumer stops purchasing at price $p$ if
\begin{equation}\label{eq:welfare_switch_support}
p<T<V\le(1+\tau)p.
\end{equation}
The lost purchase has surplus $T-p$ under TRJ and $V-p$ under RUM--APT. The latter cannot exceed the tax amount $\tau p$ because a consumer with a larger surplus would still prefer to purchase at the tax-inclusive price. This provides an upper bound on individual excess burden under RUM--APT that is robust to any joint distribution of $V$ and $T$, and requires information only about the price and tax rate. 

Consider the population of binding-threshold consumers, i.e., consumers with $C_A(0)=T<V$ for some $T$. Let $n$ be the smallest positive integer strictly greater than $p$, and let
\[
s(p,\tau):=\Prb\bigl(C_A(\tau)\le p<T\bigr)
\]
be the fraction of these consumers whose purchase is eliminated by the tax. Since thresholds are integers, every lost purchase has $T\ge n$. We will save on notation and write $s=s(p,\tau)$ when possible.

\begin{samepage}
\begin{proposition}[Bounds on mean excess burden]\label{prop:fixed_welfare}
Let $p>0$ and $\tau>0$. If $(1+\tau)p\le n$, no binding-threshold consumer changes her purchase and both mean excess burdens are zero. Otherwise,
\begin{align}
(n-p)s&\le\E[\mathrm{EB}_{\mathrm{TRJ}}(p)]\le\E[\mathrm{EB}_{\mathrm{APT}}(p)]\le\tau ps.\label{eq:mean_apt_support}
\end{align}
Whenever $s(p,\tau)>0$, the ratio of mean excess burdens satisfies
\begin{equation}\label{eq:mean_ratio_support}
1\le
\frac{\E[\mathrm{EB}_{\mathrm{APT}}(p)]}
{\E[\mathrm{EB}_{\mathrm{TRJ}}(p)]}
\le\frac{\tau p}{n-p}.
\end{equation}
\end{proposition}
\end{samepage}

The lower bound on the ratio of mean excess burdens follows from Proposition~\ref{prop:welfaredom}. For the upper bound, TRJ assigns at least $n-p$ of surplus to every lost purchase, while RUM--APT assigns at most $\tau p$. Thus, the upper bound compares the largest possible APT surplus loss with the smallest possible TRJ surplus loss among lost purchases. Consumers who do not switch contribute zero under both interpretations. If no one switches, both means are zero and their ratio is undefined. Observe that neither the switching probability nor an additional assumption on the distribution of $V$ and $T$ is needed to obtain the bounds on the ratio of mean excess burdens.

The dollar bounds are particularly simple when $n<(1+\tau)p\le n+1$. In this case, only consumers with $T=n$ can switch, so
\begin{equation}\label{eq:single_trj_exact}
\E[\mathrm{EB}_{\mathrm{TRJ}}(p)]=(n-p)s(p,\tau).
\end{equation}
If $(1+\tau)p$ is large enough so that consumers from higher thresholds can switch as well, then $(n-p)s(p,\tau)$ provides a lower bound.

For example, at $p=4.95$ and a 10-percent tax, only consumers with $T=5$ can switch. Each lost purchase has TRJ surplus of five cents and RUM--APT surplus between five and 49.5 cents. The ratio of mean losses is therefore between 1 and 9.9 whenever switching is positive. The bounds hold regardless of how valuations are distributed across thresholds.

\paragraph{Prices near an integer.}
Holding the tax rate fixed, the upper bound on the ratio of mean excess burdens $\tau p/(n-p)$ increases as price approaches $n$ from below, while the lower bound remains one. At a 10-percent tax, moving from $\$4.95$ to $\$4.99$ raises the upper bound from 9.9 to 49.9. The reason is that TRJ's surplus on a lost purchase with $T=5$ falls from five cents to one cent, while the maximum possible surplus lost from APT rises. However, a higher price means less surplus lost for consumers who would have switched even with the lower price. Thus, the mean dollar losses may go down even if the ratio of excess burdens goes up. Empirical evidence suggests that firms do not set prices optimally to account for left-digit bias and setting prices to end in $99$ cents can raise firms' profits (see \citet{ListEtAl2023} and \citet{StrulovShlain2023}). Thus, it is worth noting that prices just below integers permit especially large ratios of mean excess burdens under our two models. 

\paragraph{Larger taxes.}
At a fixed price, the upper bound on the ratio of mean excess burdens increases with the tax rate. However, a larger tax need not increase the actual ratio, even if the upper bound increases. This can occur when consumers with higher thresholds begin to switch their choice. Those consumers lose a larger TRJ surplus $T-p$, corresponding to their higher threshold $T$. Thus, the ratio may fall. If only threshold-$n$ consumers can switch, then the TRJ dollar upper bound is unchanged and the APT dollar upper bound increases proportionally with $\tau$. For a fixed population of types, a larger tax weakly increases the number of consumers who switch, so actual mean excess burden under each interpretation is also weakly increasing.

\subsection{Quantitative welfare bounds}\label{sec:welfare_quant}

We use the partially-identified overshoot distributions from Section~\ref{sec:overshoot_pi} to sharpen the theoretical bounds given earlier on the ratio of mean excess burdens. For this quantitative exercise, we add the assumption that $R=V/T$ has the same distribution across binding thresholds. 

\begin{assumption}[Common overshoots across thresholds]\label{ass:welfare_independence}
The distribution of $R=V/T$ among binding types is the same at every threshold in $\mathcal N_I$. Thus, for each $n\in\mathcal N_I$ with positive binding mass,
\[
\Prb(R\le r\mid T=n,V>n)=F_{R}(r).
\]
\end{assumption}
Equivalently, $R$ and $T$ are independent among binding types. Note that this does not imply that $V$ and $T$ are independent; in fact, it implies that the conditional distribution of willingness to pay scales proportionally with the attention threshold. 

The independence assumption connects the arm-wide evidence about overshoots (from arm-wide moments) to purchases at a fixed dollar price. For each candidate overshoot distribution satisfying all four estimated response and debunching moments, we calculate which consumers stop buying and the surplus of their purchases under each interpretation. We average using the no-tax threshold shares and obtain bounds by considering every such distribution. Appendix~\ref{app:welfare_lp} gives the details.

Table~\ref{tab:welfare_ratio} reports the results for six price and tax combinations. Purchase switching probability and dollar losses average over the 2,803 no-tax consumer-product observations classified as binding types with thresholds from 2 through 14, including nonswitchers. We do not include non-binding threshold consumers in our welfare analysis.

\begin{table}[H]
\centering
\caption{Fixed-price switching and welfare bounds}
\label{tab:welfare_ratio}
\footnotesize
\setlength{\tabcolsep}{3pt}
\begin{tabular}{rrrrrr}
\toprule
Price (\$) & Tax (\%) & Switch (\%) & APT EB (cents) & TRJ EB (cents) & APT/TRJ \\
\midrule
\multicolumn{6}{l}{\textit{Common overshoot distribution}} \\
3.95 & 6 & $[3.14,\ 6.14]$ & $[0.157,\ 0.868]$ & $[0.157,\ 0.307]$ & $[1.00,\ 2.83]$ \\
3.95 & 10 & $[5.88,\ 7.75]$ & $[0.779,\ 1.526]$ & $[0.294,\ 0.387]$ & $[2.65,\ 3.94]$ \\
3.99 & 10 & $[6.40,\ 7.80]$ & $[0.697,\ 1.263]$ & $[0.064,\ 0.078]$ & $[10.89,\ 16.19]$ \\
4.95 & 10 & $[4.78,\ 6.20]$ & $[0.752,\ 1.467]$ & $[0.239,\ 0.310]$ & $[3.15,\ 4.73]$ \\
4.95 & 30 & $[9.26,\ 16.71]$ & $[4.943,\ 16.028]$ & $[3.634,\ 5.633]$ & $[1.36,\ 3.32]$ \\
9.95 & 10 & $[4.04,\ 5.07]$ & $[1.147,\ 2.176]$ & $[0.202,\ 0.253]$ & $[5.67,\ 8.59]$ \\
\addlinespace
\bottomrule
\end{tabular}

\vspace{0.4em}
\begin{minipage}{\textwidth}\footnotesize
Notes: Switching and excess burden average over all binding threshold observations with $T\in\{2,\ldots,14\}$, using no-tax threshold shares. Switching probability corresponds to the share of binding threshold consumers who switch to not purchasing the good. Excess burden includes zeros for nonswitchers and is in cents per binding observation. APT/TRJ is the ratio of mean dollar excess burdens. We impose common overshoots across thresholds. All scenarios use the same four estimated response and debunching restrictions. Intervals are identification bounds.
\end{minipage}
\end{table}

At $\$3.99$ with a 10-percent tax, only consumers with $T=4$ can switch. Their purchase is eliminated if $R\le4.389/4=1.09725$. Thus, 42.83--52.20 percent of these types switch. This corresponds to 6.40--7.80 percent of all binding observations, the denominator used in Table~\ref{tab:welfare_ratio}. Mean RUM--APT excess burden per type with $T=4$, including nonswitchers, is 4.67--8.45 cents, compared with 0.428--0.522 cents under TRJ. The ratio of mean losses is 10.89--16.19. Thus, the four estimated moments and independence between $R$ and $T$ substantially narrow the theoretical bound of 1--39.9 obtained from Proposition~\ref{prop:fixed_welfare}.

The price comparisons illustrate why a large ratio need not imply a large dollar loss. With a 10-percent tax, moving from $\$3.95$ to $\$3.99$ raises the ratio interval from 2.65--3.94 to 10.89--16.19, even though both endpoints of the RUM--APT dollar-loss interval fall. The TRJ surplus of each lost purchase falls from five cents to one cent. At $\$4.95$, raising the tax from 10 to 30 percent raises both models' dollar-loss bounds, but lowers the upper ratio bound. The larger tax eliminates purchases by some types with $T=6$, whose TRJ surplus is $\$1.05$ rather than the five cents assigned to a purchase with $T=5$.

\section{Discussion}\label{sec:discussion}
\paragraph{Left-digit bias}
A common explanation for left-digit bias is that it arises from the consumer \emph{misperceiving} the prices she observes (see \citet{ListEtAl2023} and \citet{StrulovShlain2023}). If the posted price at the grocery store for an item is \$9.99, a left-digit biased consumer is excessively influenced by the left-most digit and perceives the price to be less than \$9.99. These are perception-based models of left-digit bias.
%Importantly, the consumer does not realize that \$9.99 is very close to \$10.

The RUM--APT model provides a different explanation for demand discontinuities at whole-dollar prices: consumers use natural attention thresholds as heuristics, and the demand discontinuity arises because a positive mass of consumers stops paying attention to the good at that price. This is a "consider then choose" explanation of left-digit bias \citep{HauserWernerfelt1990,MasatliogluEtAl2012}. A consumer may deliberately apply the attention threshold choice procedure. For example, a RUM--APT consumer can first filter goods by prices on an online retail platform, and then select her best option from the remaining options. Standard choice data, given by $Q(p)$, may not be able to separate the two explanations. 
%RUM--APT can accommodate a comparatively flat continuous portion of demand between integers through the distribution of preferences.

Distinguishing between the two explanations matters for welfare analysis. Consider a consumer who buys the good at \$9.99 but does not at \$10. In perception-based models of left-digit bias, this may occur because the consumer has a true willingness to pay of \$9.80 but misperceives the \$9.99 price to be \$9.70. When the price changes to \$10, the left-most digit changes, and she correctly perceives the new price and stops buying the good. Her error is that she buys when she should not. As a consequence, a price increase may even be beneficial to the consumer. Those same choices can be explained using our price-based attention model. The consumer has an attention threshold at \$10 and a true willingness to pay at, say, \$10.40. At \$9.99, the consumer optimally chooses the good. When the price changes to \$10, the consumer stops buying because she stops paying attention to the good. Her error is that she does not buy when she should. 

\paragraph{Robust behavioral welfare analysis}
\citet{TRJ2018}'s approach to robust behavioral welfare analysis stems from the framework outlined by \citet{BernheimRangel2009}. Broadly, the analyst identifies decision problems for which choices are informative about welfare and uses those choices to conduct welfare analysis. \citet{TRJ2018} do this by assuming that a choice with no tax is optimal, so the untaxed cutoff reflects the consumer's true willingness to pay $V$. Then, the behaviorally defined tax response, $\theta$, describes how purchases change with a tax. Combining the tax response with willingness to pay identified from untaxed cutoffs helps us learn about the lost surplus from taxation. Overall, the approach is robust to various forms of bounded rationality generating the tax response under the assumption that untaxed cutoffs reflect willingness to pay. 

%Using experiments, the researcher can create choice environments that "debias" the consumer and induce rational behavior. 

It is easy to see why excess burden with attention-price thresholds is different. With attention thresholds, choices both with and without taxes are suspect. A mass of integer cutoffs reflects attention almost surely. Taxes induce debunching from that mass, and we interpret a tax-induced non-integer cutoff as an undistorted reflection of the consumer's posted-price willingness to pay, $V/(1+\tau)$. From this optimal choice, we can then recover her posted-price willingness to pay with no tax, $V$. For any binding-threshold consumer whose purchase is eliminated by the tax, willingness to pay exceeds her attention threshold, so excess burden is greater than under TRJ.\footnote{Note that even with a tax, the remaining mass at thresholds can behave sub-optimally. Nevertheless, excess burden of taxation is identified from individual cutoffs because we do not need to know the willingness to pay of consumers who never change their choice. } 

Differences in mechanisms are relevant even when models agree on some measures of welfare. Consider a 10-percent tax and take a binding-threshold consumer with the same tax and no-tax cutoff of $\$10$. This person would choose a good priced $\$9.99$ and both models would agree that the excess burden of tax is zero. Under the TRJ interpretation, with a tax she is choosing the good when she shouldn't. She would be better off if she did not choose the good. The tax creates a private welfare loss since now she has to give a transfer to the government. In our model, her choice with a tax is optimal. By not choosing, the consumer would be worse off. 

So, which choices reveal preferences? There are three reasons to consider the attention threshold interpretation. First, the standard tax module reflects the environment the consumer faces when making choices in real life. The tax is equal to the consumer's local sales tax, and when consumers make purchases, this tax is not included in the posted price. Thus, if consumers are following a heuristic-based choice rule, then that setting is where their choices are more likely to be correct, and the unfamiliar no-tax setting is where they may make errors. Second, consumer surveys from \citet{TRJ2018}, \citet{FeldmanRuffle2015} and \citet{CLK2009} find that consumers are well-aware of their sales tax. \citet{FeldmanRuffle2015} and \citet{TRJ2018} (in Appendix E.2) find that consumers are able to compute the tax correctly as well. In addition, \citet{CLK2009} find that demand returns to pre-experiment levels after they stop their tax-inclusive posted price intervention. These findings are consistent with attention-thresholds where the tax is processed fully, but we still get under-reaction. Third, the left-digit bias literature provides further reason to question whether no-tax choices necessarily reflect undistorted preferences.

\clearpage
\appendix
\section{Proofs}\label{app:proofs}

\subsection{Proof of Proposition \ref{propnotaxid}}

\begin{proof}
Let $n\in\N_+$. Since the choice probability $Q(p)=\Prb(C>p)$, the observed integer discontinuity
\[
J_n=Q(n-)-Q(n)=\Prb(C=n).
\]
Since $C=\min\{T,V\}$ and $T\in\N_+\cup\{\infty\}$, we have that $C=n$ if and only if $T=n$ and $V\ge n$, or  $V=n$ and $T\ge n$. By Assumption~\ref{assnoties}, $\Prb(V=n)=0$. Therefore
\[
\Prb(C=n)=\Prb(T=n,V>n).
\]
Thus the observed discontinuity $J_n=Q(n-)-Q(n)$ identifies the binding threshold mass $\Prb(T=n,V>n)$.

\end{proof}

\subsection{Proof of Proposition \ref{prop:debunch}: tax-induced debunching}

\begin{proof}
Let $n\in\N_+$. We know that
\[
J_n(\tau)=Q_\tau(n-)-Q_\tau(n)
\]
is the total mass at cutoff $n$ under tax $\tau$. Assumption~\ref{ass:continuousnoatoms} implies that the continuous residual contributes no atom at $n$. Thus, any atom at $n$ must come from binding threshold types. Binding-threshold types whose taxed cutoff equals $n$ can be divided into two groups: those whose no-tax cutoff was also $n$ and those whose no-tax cutoff was not $n$. A positive mass can remain at $n$ with tax only when $T=n$ and $V/(1+\tau)\ge n$. Any type with $T>n$ would require $V/(1+\tau)=n$ to have cutoff equal to $n$, which has probability zero by Assumption~\ref{assnoties}. Finally, $T<n$ cannot generate cutoff at $n$. Therefore
\[
J_n(\tau)
=
\Prb\!\left(T=n,\;V\ge n(1+\tau)\right)
=
\Prb\!\left(T=n,\;V>n(1+\tau)\right),
\]
which proves equation~\eqref{eq:Jtax_structural}. As $\tau$ increases,  $\Prb\!\left(T=n,\;V>n(1+\tau)\right)$ weakly decreases. Thus, $J_n(\tau)$ is weakly decreasing in the tax rate.

Now suppose $J_n(0)>0$. Using equation~\eqref{eq:Jtax_structural},
\[
\frac{J_n(\tau)}{J_n(0)}
=
\Prb\!\left(V>n(1+\tau)\mid T=n,V>n\right)
=
\Prb\!\left(\frac{V}{n}>1+\tau\mid T=n,V>n\right).
\]
Taking complements gives equation~\eqref{eq:debunchcdf}. 
\end{proof}

\subsection{Proof of Proposition \ref{prop:nonewatoms}: no new atoms under taxation}

\begin{proof}
Let $\tau>0$ and let $p>0$ be a noninteger price. By Assumption~\ref{ass:continuousnoatoms}, the continuous residual contributes no atom at $p$ with a tax. Therefore, any atom at $p$ must come from binding threshold types, whose taxed cutoff is
\[
C_A(\tau)=\min\left\{T,\frac{V}{1+\tau}\right\}.
\]
Since $p\notin\N_+$, the threshold itself cannot generate a cutoff at $p$ because $T$ is integer-valued. Hence, a binding threshold type can have a cutoff at $p$ only if
\[
\frac{V}{1+\tau}=p.
\]
By Assumption~\ref{assnoties} (no utility ties), this occurs with probability zero. Thus, the full taxed cutoff distribution has no atom at any noninteger $p$.

Finally, consider a binding threshold-$n$ type who moves under the tax. She moves if and only if $n<V<n(1+\tau)$. Her new cutoff with the tax is
\[
C_A(\tau)=\frac{V}{1+\tau}.
\]
Therefore,
\[
\frac{n}{1+\tau}<C_A(\tau)<n,
\]
which gives equation~\eqref{eq:moversupport}. For any price $p$ in this interval, the cutoff $C_A(\tau)=p$ if and only if $V=p(1+\tau)$. By Assumption~\ref{assnoties} this occurs with probability zero, and conditioning on a positive-probability mover event cannot create an atom that is absent unconditionally. Hence the mover cutoff distribution is atomless.
\end{proof}

\subsection{Proof of Proposition \ref{prop:thresholdresponse}: threshold tax response}

\begin{proof}
Consider a binding threshold type with $C_A(0)=T$ and $R=V/T>1$. If $R\geq1+\tau$, then
\[
\frac{V}{1+\tau}=\frac{RT}{1+\tau}\geq T,
\]
so the attention threshold continues to bind and $C_A(\tau)=T$. Her tax response is $\psi_A(\tau)=0$.

If $R<1+\tau$, then the attention threshold no longer binds and
\[
C_A(\tau)=\frac{V}{1+\tau}=\frac{RT}{1+\tau}.
\]
Her tax response is
\[
\psi_A(\tau)
=
\frac{(1+\tau)/R-1}{\tau}
=
\frac{1+\tau-R}{R\tau}.
\]
Combining the two cases proves equation~\eqref{eq:psiA_formula}. Because $R>1$, the tax response is nonnegative and strictly below one whenever it is positive.

For $\tau\leq R-1$, $\psi_A(\tau)=0$. For $\tau>R-1$,
\[
\frac{\mathrm{d} \psi_A(\tau)}{\mathrm{d} \tau}
=
\frac{R-1}{R\tau^2}>0.
\]
The function is continuous at $\tau=R-1$, where it equals zero. Thus $\psi_A(\tau)$ is weakly increasing in $\tau$ and strictly increasing after the threshold ceases to bind. The population result follows by integrating the pointwise inequality with respect to any fixed distribution of $R$ in the binding-threshold population.
\end{proof}

\subsection{Proof of Proposition \ref{prop:paired}}

\begin{proof}
We know that the no-tax cutoff $C=\min\{T,V\}$. By Assumption~\ref{assthresholdsupport}, every finite $T$ is an integer, so a noninteger value of $C$ cannot be generated by a binding finite threshold. Hence, if $C$ is noninteger, then $V=C$. Now let $C=n\in\N_+$. By Assumption~\ref{assnoties}, $V=n$ has probability zero. Therefore, the no-tax integer cutoff is generated by a binding threshold, $T=n<V$, almost surely. If $C_A(\tau)<n$, the threshold no longer binds under tax, so equation \eqref{eq:thresholdcutoff} implies $C_A(\tau)=V/(1+\tau)$ and therefore $V=(1+\tau)C_A(\tau)$. If $C_A(\tau)=n$, the threshold continues to bind, which requires $V/(1+\tau)\ge n$, or $V\ge n(1+\tau)$.
\end{proof}

\subsection{Proof of Proposition \ref{prop:partialover}}

\begin{proof}
Fix a tax rate $\tau\in\Lambda$. By Proposition~\ref{prop:debunch},
\[
J_n(0)-J_n(\tau)
=
\Prb(T=n,V>n,R\le1+\tau).
\]
Summing over all the attention thresholds and dividing
by the total binding-threshold mass gives
\[
1-\frac{\sum_{n\in\mathbb N_+}J_n(\tau)}
        {\sum_{n\in\mathbb N_+}J_n(0)}
=
\Prb(R\le1+\tau\mid T<V).
\]
For the binding-threshold population, $Z=\log R$
has CDF $F_Z$. Since the logarithm is strictly increasing, it is a one-to-one transformation of $R$, so we get 
\[
\Prb(R\le1+\tau\mid T<V)
=
F_Z\!\left(\log(1+\tau)\right).
\]
Thus, the aggregate debunching probability identifies
the share of consumers with valuation overshoot no greater than $1+\tau$.
Let $a=\log(1+\tau)$. All expectations below are taken within the population of binding-threshold consumers. Equation~\eqref{eq:phib_formula} implies
\[
\tau\E[\phi_A(\tau)]=\E[(a-Z)_+].
\]
Applying the layer-cake identity gives
\[
(a-Z)_+=\int_0^a \1\{Z\le z\}\dd z.
\]
Then, taking expectations and applying Tonelli's theorem yields
\[
\E[(a-Z)_+]=\int_0^a F_Z(z)\dd z,
\]
so the mean log threshold response, with the known tax rate, identifies the stated integral. The argument applies to every $\tau\in\Lambda$. The above equations give the moments generated by any overshoot distribution satisfying the model assumptions. Such a distribution reproduces the observed moments if and only if its CDF satisfies the above equalities at every $\tau\in\Lambda$.
\end{proof}

%\subsection{Proof of Proposition \ref{prop:trjrepresentation}}
%
%\begin{proof}
%Fix a tax rate $\tau>0$. For a binding-threshold mover, the no-tax cutoff $C_A(0)=T<V$ and the taxed cutoff
%\[
%C_A(\tau)=\frac{V}{1+\tau}<C_A(0).
%\]
%Since $V>C_A(0)$,
%\[
%C_A(\tau)=\frac{V}{1+\tau}>\frac{C_A(0)}{1+\tau}.
%\]
%Combining the two inequalities gives
%\[
%1<\frac{C_A(0)}{C_A(\tau)}<1+\tau.
%\]
%Then, subtracting one and dividing by $\tau>0$ gives $0<\theta^*<1$. For a binding-threshold stayer, we have that the no-tax and taxed cutoffs are the same, so $\theta^*=0$. Finally, construct the TRJ representation by choosing $V^{\mathrm{TRJ}}$ and $\theta^*$ according to equations~\eqref{eq:trj_wtp} and ~\eqref{eq:trj_theta}, respectively. By construction,
%\[
%\frac{V^{\mathrm{TRJ}}}{1+\theta^*\tau}
%=
%\frac{C_A(0)}{1+\bigl(C_A(0)/C_A(\tau)-1\bigr)}
%=C_A(\tau),
%\]
%while the no-tax cutoff is $V^{\mathrm{TRJ}}=C_A(0)$.
%\end{proof}

\subsection{Proof of Proposition \ref{prop:welfaredom}}

\begin{proof}
Consider a binding-threshold consumer. If her purchase is eliminated by the tax, by equation~\eqref{eq:welfare_gap} we know that 
\[
\mathrm{EB}_{\mathrm{APT}}(p)-\mathrm{EB}_{\mathrm{TRJ}}(p)
=V-T>0.
\]
If the consumer does not change her choice, then the excess burden, under both models, is zero. Therefore, $\mathrm{EB}_{\mathrm{APT}}(p)\ge\mathrm{EB}_{\mathrm{TRJ}}(p)$ for every binding-threshold consumer. This implies that, for every $z$, $G_{\mathrm{APT}}(z)\le G_{\mathrm{TRJ}}(z)$. If a positive mass of binding-threshold consumers have their purchase eliminated by the tax at $p$, then $\mathrm{EB}_{\mathrm{APT}}(p)>\mathrm{EB}_{\mathrm{TRJ}}(p)$ with positive probability. Thus, the first-order stochastic dominance is strict for at least one $z$.
\end{proof}

\subsection{Proof of Proposition~\ref{prop:fixed_welfare} }\label{app:fixed_welfare_proofs}

\begin{proof}
For a binding type, $C_A(0)=T$ and $C_A(\tau)=\min\{T,V/(1+\tau)\}$. Hence, the purchase is eliminated when $p<T<V\le(1+\tau)p$. Since $T$ is an integer and $n$ is the smallest integer above $p$, every switcher has $T\ge n$. If $(1+\tau)p\le n$, switching is impossible. If the consumer switches, the following must be true:
\[
n-p\le T-p<V-p\le\tau p.
\]
Recall that $\mathrm{EB}_{\mathrm{TRJ}}(p)=T-p$ and $\mathrm{EB}_{\mathrm{APT}}(p)=V-p$. If the consumer does not switch, then excess burdens under both models are zero. 

Finally, we consider expected excess burden in the binding-threshold population. Consumers who do not switch contribute zero to the excess burden. The share who do switch, $s(p,\tau)$, contribute the excess burden given above for each model. Taking the expectation, we get equation~\eqref{eq:mean_apt_support}. If $s(p,\tau)>0$, then $\E[\mathrm{EB}_{\mathrm{TRJ}}(p)]\ge(n-p)s(p,\tau)>0$. Dividing by the mean excess burden under TRJ gives equation~\eqref{eq:mean_ratio_support}.
\end{proof}

\clearpage
\section{Additional empirical details}\label{app:empirical}

\subsection{Sample construction and binding-threshold observations}\label{app:sample_cells}

The empirical cutoffs are the TRJ replication variables \texttt{up1} and \texttt{up2}, denoted $P^1$ and $P^2$ in the paper. Consumers use a slider to report the maximum posted price they are willing to pay for each product. These cutoffs equal the slider reports unless a report reaches the upper endpoint of \$15, in which case they incorporate the consumer's follow-up response about the maximum tag price she is willing to pay.

Following TRJ's replication code, we retain consumers who report no zero cutoffs for any product in either module (\texttt{zero=0}), have no missing follow-up responses needed to construct their cutoffs (\texttt{missing=0}), face a positive local sales-tax rate (\texttt{taxc>0}), and correctly answer the primary questions about the tax treatment in Module 1 (\texttt{pass=1}). In the public replication file, the first three restrictions remove no observations. The comprehension restriction reduces the sample from 4,223 to 2,998 consumers. We then retain consumer-product observations with an untaxed Module 2 cutoff of at least one dollar, leaving 58,478 observations from 2,991 consumers. Of these, 21,551 observations from 1,098 consumers belong to the no-tax arm, 19,100 observations from 979 consumers to the standard-tax arm, and 17,827 observations from 914 consumers to the triple-tax arm.

\begin{figure}[H]
\centering
\includegraphics[width=0.82\textwidth]{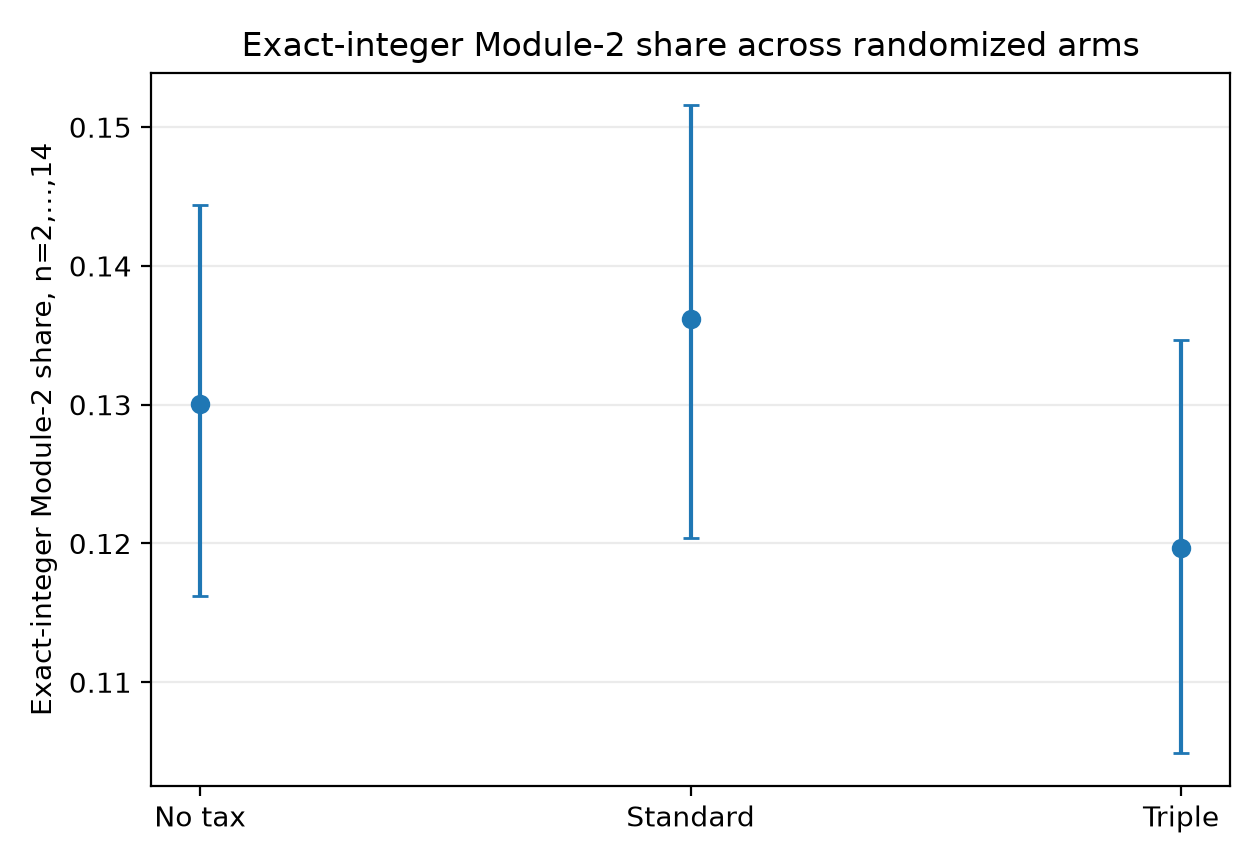}
\caption{Share of integer reports in Module 2 across arms. Error bars show 95-percent confidence intervals given by the 2.5th and 97.5th percentiles of 5,000 consumer-cluster bootstrap estimates.}
\label{fig:arm}
\end{figure}

We identify binding-threshold observations using exact integer reports from 2 through 14 in untaxed Module 2. Figure~\ref{fig:arm} reports the fraction of binding-threshold observations in each experimental arm.

\subsection{Descriptive Statistics}\label{app:desc stats}
Table~\ref{tab:binding_demographics} reports demographic characteristics of consumers with at least one binding-threshold observation. We find no significant difference across the three arms for most of the demographic variables. However, postgraduate education differs significantly across arms $(p\approx0.01)$.

\begin{table}[H]
\centering
\caption{Demographics of consumers with binding-threshold observations}
\label{tab:binding_demographics}
\footnotesize
\setlength{\tabcolsep}{3pt}
\begin{tabular}{lrrrrr}
\toprule
 & All & No tax & Standard & Triple & $F$-test $p$-value \\
\midrule
Age & 50.33 & 50.73 & 50.37 & 49.73 & 0.59 \\
 & (14.69) & (14.36) & (14.71) & (15.11) & \\
Household income (\$1,000s) & 60.52 & 60.86 & 64.12 & 55.92 & 0.09 \\
 & (54.84) & (51.83) & (58.00) & (54.88) & \\
Household size & 2.42 & 2.43 & 2.43 & 2.40 & 0.92 \\
 & (1.56) & (1.67) & (1.55) & (1.42) & \\
Married & 0.60 & 0.61 & 0.62 & 0.57 & 0.29 \\
Male & 0.44 & 0.42 & 0.47 & 0.43 & 0.29 \\
\addlinespace
\multicolumn{6}{l}{\textit{Education}} \\
High school degree or higher & 0.95 & 0.95 & 0.95 & 0.95 & 0.99 \\
College degree or higher & 0.38 & 0.40 & 0.35 & 0.39 & 0.22 \\
Postgraduate education & 0.14 & 0.17 & 0.11 & 0.13 & 0.01 \\
\addlinespace
\multicolumn{6}{l}{\textit{Ethnicity}} \\
Asian & 0.03 & 0.03 & 0.04 & 0.03 & 0.67 \\
Caucasian & 0.76 & 0.75 & 0.78 & 0.76 & 0.61 \\
Hispanic & 0.04 & 0.04 & 0.03 & 0.03 & 0.50 \\
African American & 0.08 & 0.08 & 0.06 & 0.10 & 0.12 \\
Other & 0.03 & 0.03 & 0.02 & 0.02 & 0.32 \\
Local sales-tax rate (\%) & 7.34 & 7.37 & 7.34 & 7.32 & 0.73 \\
 & (1.15) & (1.16) & (1.14) & (1.14) & \\
\addlinespace
Consumers & 1,450 & 562 & 474 & 414 & \\
Binding-threshold observations & 7,537 & 2,803 & 2,601 & 2,133 & \\
\bottomrule
\end{tabular}

\vspace{0.4em}
\begin{minipage}{\textwidth}\footnotesize
Notes: The sample includes consumers in the estimation sample who report at least one untaxed Module 2 cutoff at an integer from 2 through 14. Entries are consumer means, with standard deviations in parentheses. Binary characteristics are reported as shares. The final column tests equality of means across the three arms using an OLS regression on arm indicators with robust standard errors. All included consumers pass the primary tax-comprehension questions.
\end{minipage}
\end{table}

\subsection{Tax response}\label{app:response_decomp}
\begin{figure}[H]
\centering
\includegraphics[width=0.80\textwidth]{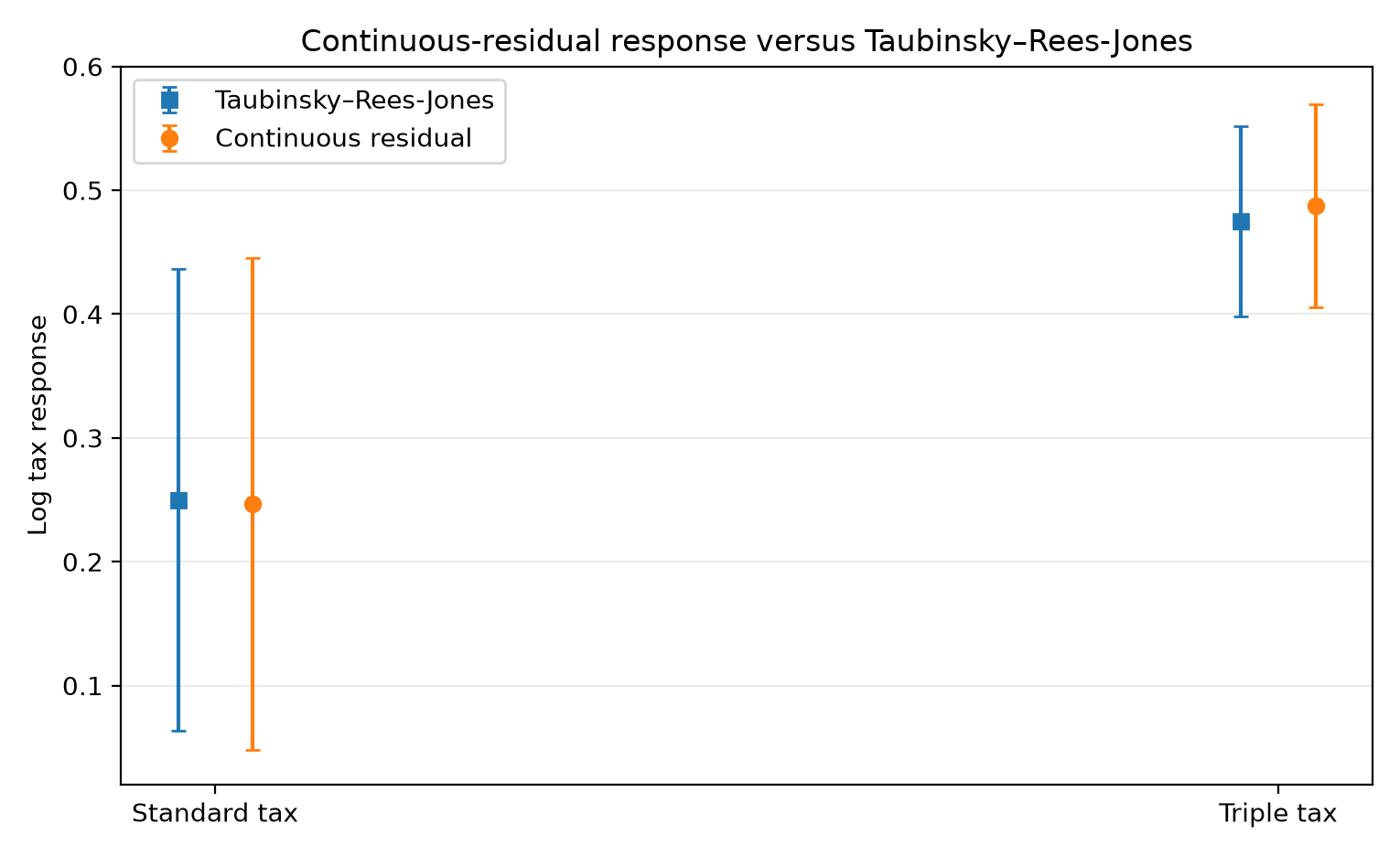}
\caption{Continuous-residual log tax responses compared with estimates using the Taubinsky and Rees-Jones baseline specification. Error bars are 95-percent normal-approximation intervals using consumer-cluster standard errors.}
\label{fig:theta}
\end{figure}

Figure~\ref{fig:theta} compares our estimated tax responses for the continuous residual with estimates obtained by applying the baseline specification from \citet{TRJ2018} to our full estimation sample in the standard- and triple-tax arms. The specification from TRJ estimates the mean tax response for the whole arm, including binding-threshold observations. 

The following figures plot the distributions of reported cutoffs across both modules and three experimental arms. The large spikes at integer prices suggest atoms in the distribution of cutoffs. 
\begin{figure}[H]
\centering
\includegraphics[width=\textwidth]{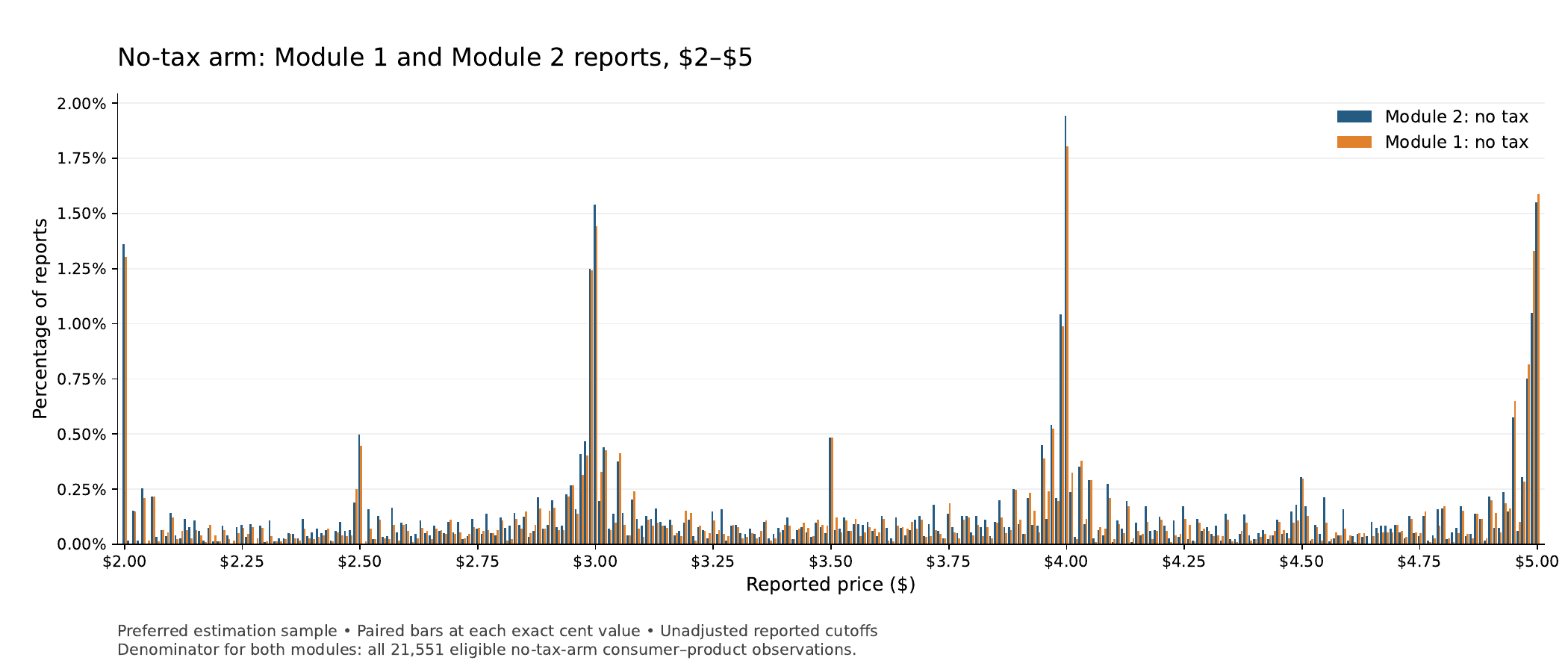}
\caption{Module 1 and Module 2 report shares from \$2 to \$5 in the no-tax arm. Blue bars show Module 2 reports; orange bars show Module 1 reports. Both modules are untaxed. Each pair of bars corresponds to one exact cent value. Both series use all 21,551 eligible no-tax arm consumer-product observations as the denominator, including reports outside the displayed interval.}
\label{fig:report_shares_2_to_5}
\end{figure}

\begin{figure}[H]
\centering
\includegraphics[width=\textwidth]{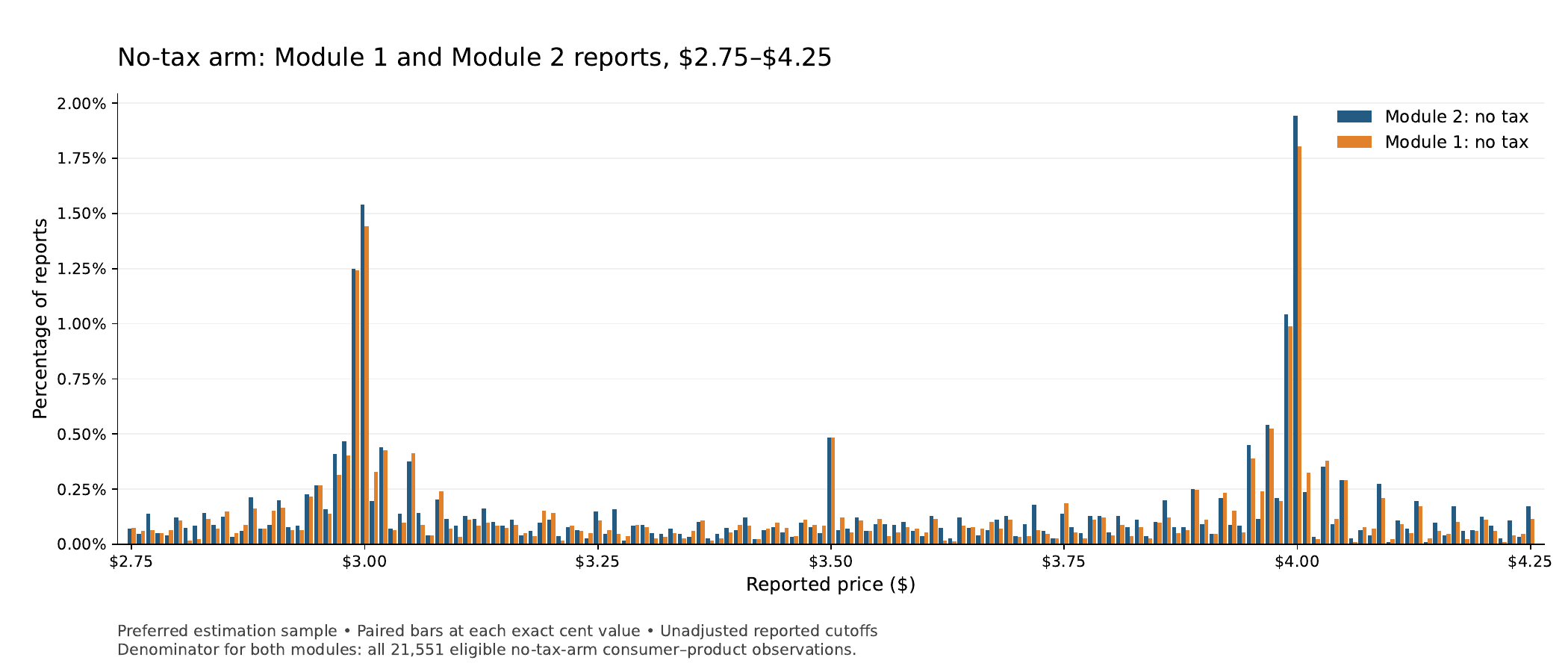}
\caption{Module 1 and Module 2 report shares from \$2.75 to \$4.25 in the no-tax arm. Blue bars show Module 2 reports; orange bars show Module 1 reports. Both modules are untaxed. Each pair of bars corresponds to one exact cent value. Both series use all 21,551 eligible no-tax arm consumer-product observations as the denominator, including reports outside the displayed interval.}
\label{fig:notax_arm_reports_2_75_to_4_25}
\end{figure}

\begin{figure}[H]
\centering
\includegraphics[width=\textwidth]{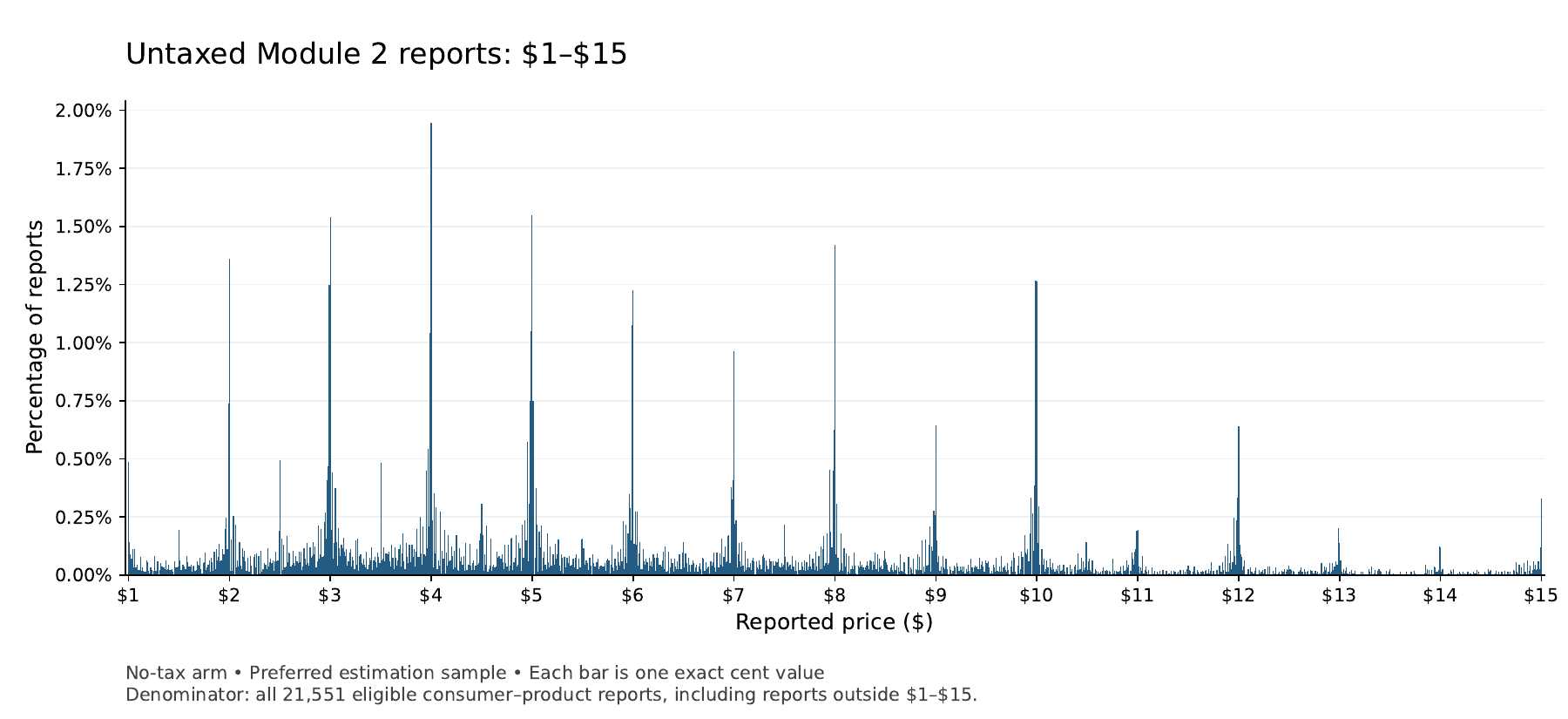}
\caption{Untaxed Module 2 report shares from \$1 to \$15 in the no-tax arm. Each bar gives the percentage of all 21,551 eligible consumer-product reports at that exact cent value. The denominator includes reports outside the displayed interval.}
\label{fig:report_shares_1_to_15}
\end{figure}

\begin{figure}[H]
\centering
\includegraphics[width=\textwidth]{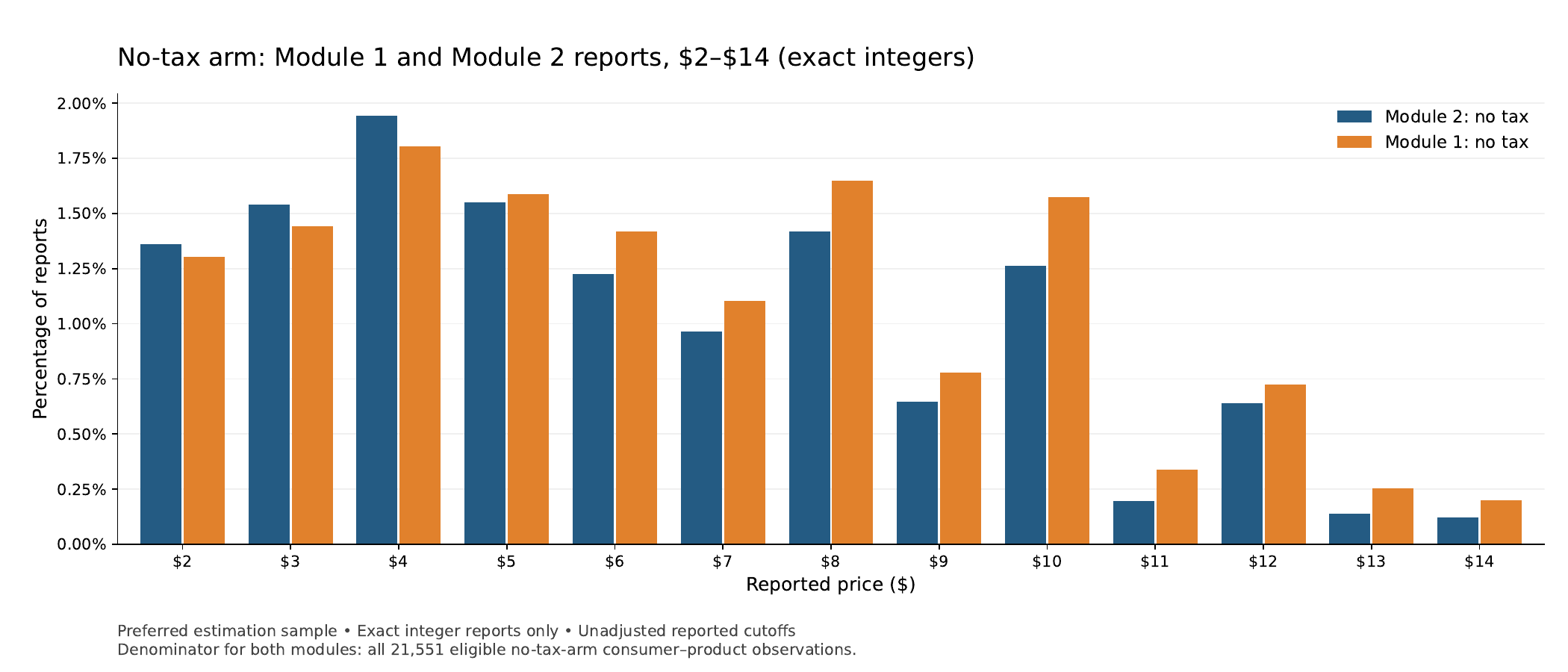}
\caption{Exact-integer report shares from \$2 to \$14 in the no-tax arm. Blue bars show Module 2 reports; orange bars show Module 1 reports. Both modules are untaxed. Each bar gives the percentage of all 21,551 eligible no-tax consumer-product observations reporting that exact integer. The reports use the replication variables \texttt{up1} and \texttt{up2}, including follow-up responses, without order-effect adjustments. The denominator includes noninteger reports and reports outside the displayed range.}
\label{fig:notax_integer_reports_2_to_14}
\end{figure}

\begin{figure}[H]
\centering
\includegraphics[width=\textwidth]{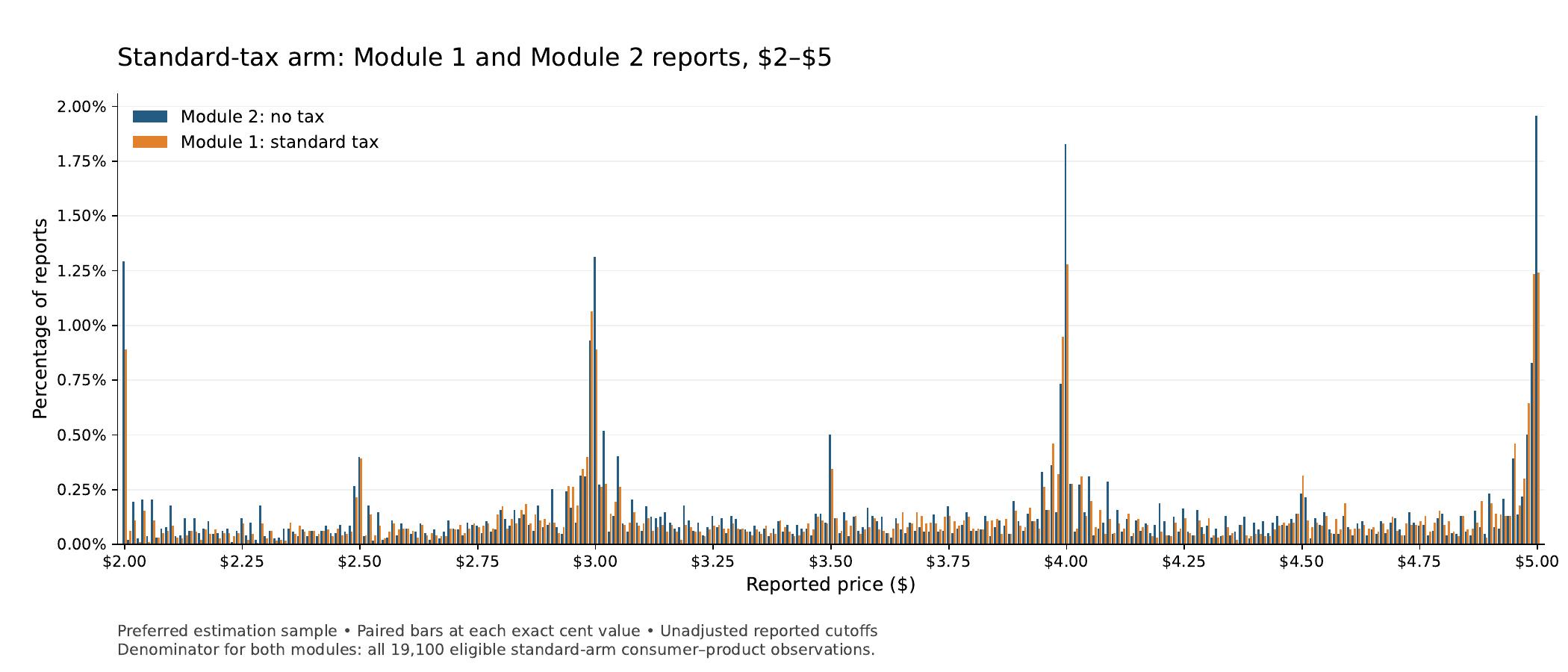}
\caption{Module 1 and Module 2 report shares from \$2 to \$5 in the standard tax arm. Blue bars show untaxed Module 2 reports; orange bars show taxed Module 1 reports. Each pair of bars corresponds to one exact cent value. Both series use all 19,100 eligible standard arm consumer-product observations as the denominator, including reports outside the displayed interval.}
\label{fig:standard_arm_reports_2_to_5}
\end{figure}

\begin{figure}[H]
\centering
\includegraphics[width=\textwidth]{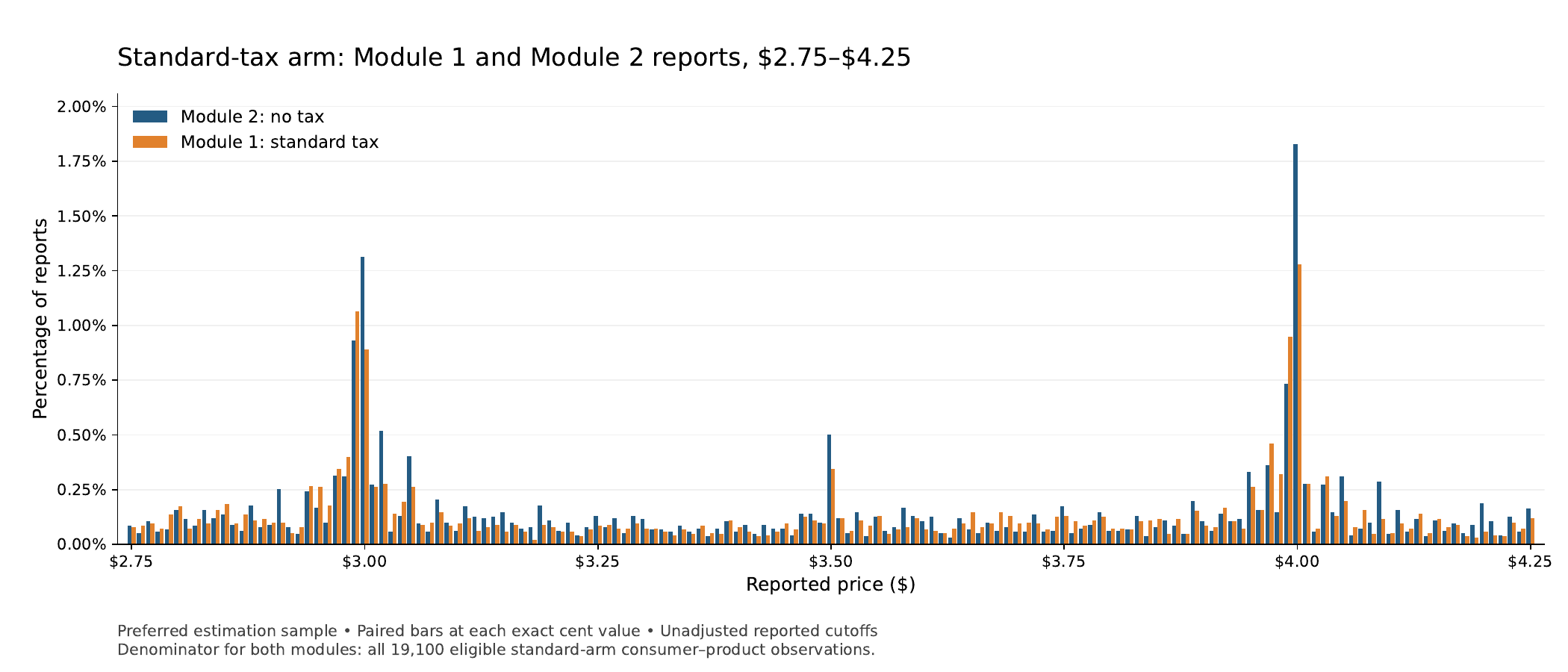}
\caption{Module 1 and Module 2 report shares from \$2.75 to \$4.25 in the standard tax arm. Blue bars show untaxed Module 2 reports; orange bars show taxed Module 1 reports. Each pair of bars corresponds to one exact cent value. Both series use all 19,100 eligible standard arm consumer-product observations as the denominator, including reports outside the displayed interval.}
\label{fig:standard_arm_reports_2_75_to_4_25}
\end{figure}

\begin{figure}[H]
\centering
\includegraphics[width=\textwidth]{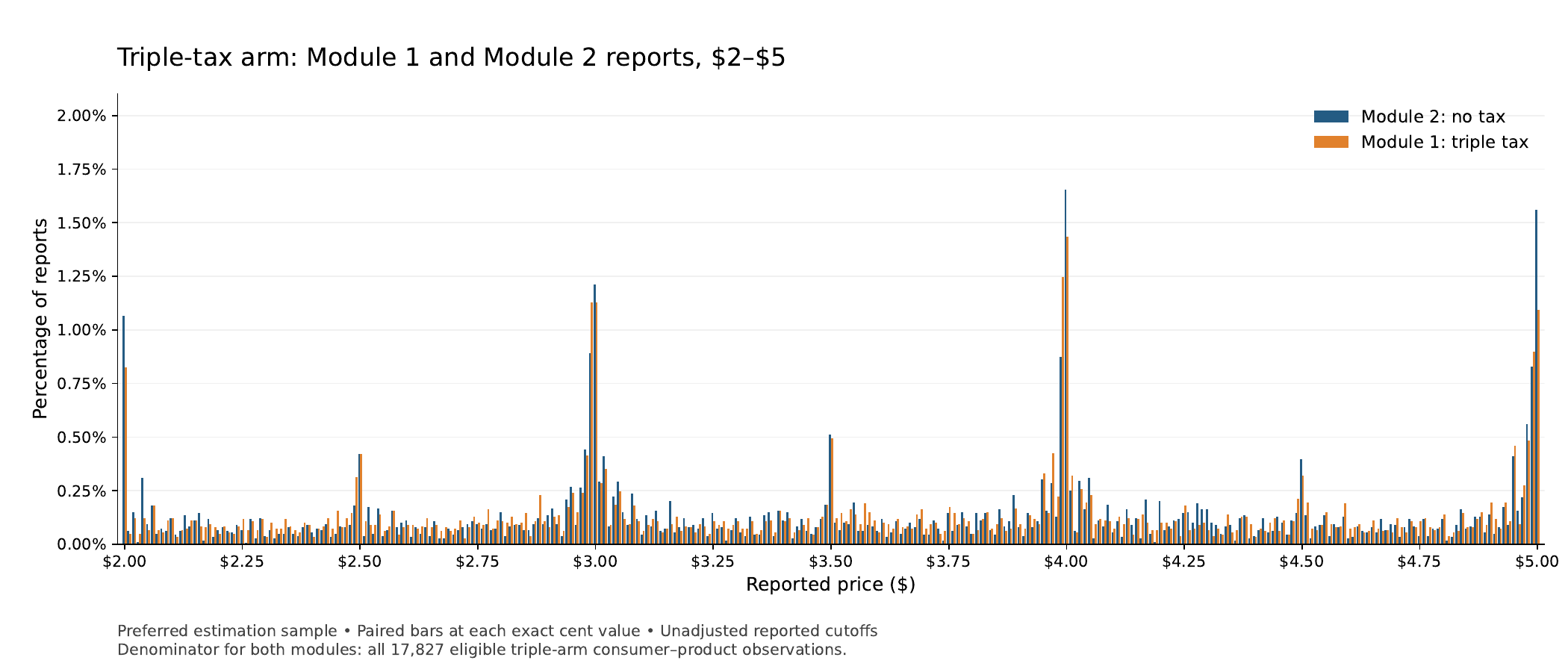}
\caption{Module 1 and Module 2 report shares from \$2 to \$5 in the triple tax arm. Blue bars show untaxed Module 2 reports; orange bars show taxed Module 1 reports. Each pair of bars corresponds to one exact cent value. Both series use all 17,827 eligible triple arm consumer-product observations as the denominator, including reports outside the displayed interval.}
\label{fig:triple_arm_reports_2_to_5}
\end{figure}

\begin{figure}[H]
\centering
\includegraphics[width=\textwidth]{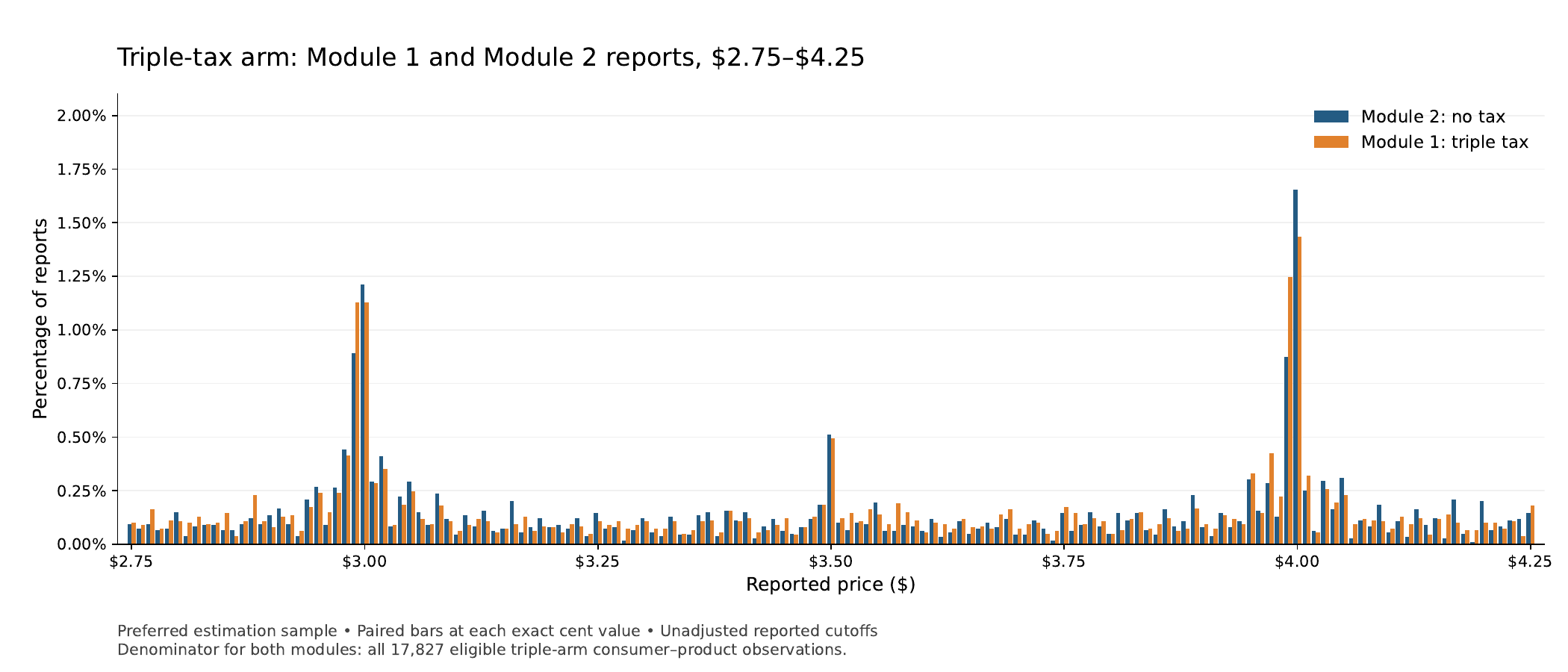}
\caption{Module 1 and Module 2 report shares from \$2.75 to \$4.25 in the triple tax arm. Blue bars show untaxed Module 2 reports; orange bars show taxed Module 1 reports. Each pair of bars corresponds to one exact cent value. Both series use all 17,827 eligible triple arm consumer-product observations as the denominator, including reports outside the displayed interval.}
\label{fig:triple_arm_reports_2_75_to_4_25}
\end{figure}

\subsection{Threshold persistence, gross debunching, and net integer mass loss}\label{app:source_deheap}

For each arm $a\in\{0,S,H\}$ and threshold $n\in\mathcal N_I=\{2,\ldots,14\}$, let $N_{a,n}$ be the number of observations with $P^2=n$, and let $K_{a,n}$ be the number with both $P^2=n$ and $P^1=n$. The estimated persistence rate and tax arm threshold weight are
\[
\widehat q_{a,n}=\frac{K_{a,n}}{N_{a,n}},
\qquad
\widehat w_{a,n}=\frac{N_{a,n}}{\sum_{m\in\mathcal N_I}N_{a,m}}.
\]
These are then used to estimate gross debunching using equation~\eqref{eq:grossdeheapagg}. 
The aggregate same-$n$ persistence rate in arm $a$ is
\[
\overline q_a=\frac{\sum_n K_{a,n}}{\sum_n N_{a,n}}.
\]
The reported persistence decline is $\overline q_0-\overline q_a$. It equals 0.137579 in the standard arm and 0.172326 in the triple arm. 

For net integer-mass loss, let $\widehat b_a$ and $\widehat u_a$ denote the fractions of all arm $a$ observations with integer Module-2 and Module-1 reports in $\mathcal N_I$, respectively. Our estimate of \textit{net integer-mass loss} is
\[
\frac{\widehat b_a-\widehat u_a}{\widehat b_a}
-\frac{\widehat b_0-\widehat u_0}{\widehat b_0}.
\]
The no-tax mass change is $-0.090260$, which means that integer mass is higher in Module 1 than in Module 2. Subtracting this from the treated changes gives 0.314789 in the standard arm and 0.251067 in the triple arm.

In five of the 5,000 consumer-cluster bootstrap draws, the no-tax persistence rate at threshold $n=13$ is zero, while that threshold has positive weight in both taxed arms. The gross debunching estimates are therefore undefined in those draws. We calculate gross debunching standard errors using the remaining 4,995 draws. Net integer-mass loss and persistence-decline estimates are defined in all 5,000 draws.

\subsection{Noninteger destination scan}\label{app:noninteger_scan}
We check whether taxation generates excess mass at a noninteger reported price. The scan considers all 1,338 noninteger cent prices from \$1.25 through \$14.75. At each candidate price $x$, we compare the share of Module 1 reports at $x$ with a benchmark constructed from neighboring prices, and then compare this excess share between the taxed and no-tax arms. The share of reports at each price is calculated using the binding-threshold observations in that arm. 

The neighboring price benchmark is constructed as follows. For each candidate price $x$, we consider the 51 cent bins in $[x-0.25,x+0.25]$, while omitting the candidate bin and every integer bin. We fit a weighted quadratic to the share of reports at the remaining bins using triangular weights. The triangular weights are $1-|r|/25$, where $r$ is a bin's distance from $x$ in cents. Thus, bins closer to $x$ receive greater weight. The fitted value at $x$ provides the neighboring price benchmark.

Let $\widehat B_a(x)$ denote the observed share of reports at $x$ minus the  share predicted by the quadratic fit in arm $a$. The difference between excess shares at price $x$ in the taxed arm and no-tax arm is
\begin{equation}\label{eq}
\widehat\Delta_a(x):=\widehat B_a(x)-\widehat B_0(x),
\end{equation}
where $a\in\{S,H\}$ denotes a taxed arm and $0$ denotes the no-tax arm. A positive difference indicates that local excess mass is larger in the taxed arm.

Standard errors are clustered at the consumer level to allow reports from the same consumer to be correlated. These account for sampling variation in the observed cutoff report shares and the fitted neighboring price benchmarks, including variation in the number of binding-threshold observations used to calculate the shares. To account for multiple hypothesis testing (since we are scanning many prices), we use a one-sided Gaussian multiplier max-$t$ procedure with 10,000 draws, adjusting jointly across all candidate prices within each treated arm. 

\subsection{Common overshoot identification}\label{app:commonovershoot}

We describe in detail the procedure used to partially identify the common overshoot distribution using the mean tax response and debunching moments. Let $\mathcal A_a$ denote the set of binding-threshold observations in treated arm $a\in\{S,H\}$, and let $N_a$ be its size. Suppose all binding-threshold observations in the arm had valuation overshoot $r$. Then, the arm-wide mean log tax response and gross debunching probability would be given by 
\[
\overline\phi_a(r):=\frac{1}{N_a}\sum_{(i,k)\in\mathcal A_a}\phi_A(\tau_i,r),
\qquad
\overline d_a(r):=\frac{1}{N_a}\sum_{(i,k)\in\mathcal A_a}\1\{r<1+\tau_i\}.
\]
The individual log tax response $\phi_A(\tau_i,r)$ is defined in equation~\eqref{eq:overshoot_functions}. We use each observation's actual tax rate, so there can be heterogeneity in the tax response and debunching even with a common overshoot $r$. 

Under the common-distribution specification, the overshoot distribution $F_R$ is the same across experimental arms and local sales-tax rates. Any overshoot distribution consistent with the data must match the four estimated arm-wide moments: two mean log tax response and two gross debunching moments. Thus, any feasible overshoot distribution satisfies
\[
\widehat m_{A,a}
=\int\overline\phi_a(r)\,\mathrm dF_R(r),
\qquad
\widehat d^{\mathrm{gross}}_a
=\int\overline d_a(r)\,\mathrm dF_R(r),
\qquad a\in\{S,H\}.
\]
The integrals average the predicted arm-wide tax response and gross debunching over candidate overshoots according to $F_R$. The four estimated moments, in the order standard-arm tax response, standard-arm gross debunching, triple-arm tax response, and triple-arm gross debunching, are
\[
(0.26853942,\ 0.41482744,\ 0.38287720,\ 0.52202521).
\]
We obtain bounds on $F_R(r)$ by minimizing and maximizing its value over distributions satisfying these four restrictions.

We compute the bounds by linear programming, representing candidate log overshoot distributions as mixtures of uniform distributions over narrow intervals. We check numerical accuracy by comparing the resulting extrema with outer bounds for the continuous distribution problem and repeating the calculation on a coarser grid.

\subsection{Quantitative welfare bounds construction}\label{app:welfare_lp}

We use the partially identified overshoot distributions in Section~\ref{sec:overshoot_pi} to calculate the welfare bounds in Section~\ref{sec:welfare_quant}. Assumption~\ref{ass:welfare_independence} ensures that each candidate overshoot distribution applies at every binding threshold. Suppose supply is perfectly elastic, the price of a good is $p$, and a tax is introduced with rate $\tau$. Given the partially identified overshoot distributions, how does the excess burden differ under the attention threshold model and the TRJ welfare interpretation? 

Let $\widehat w_{0,n}$ denote the fraction of the binding-threshold observations in the no tax arm with Module 2 report $n$. For a posted price $p$ and tax rate $\tau$, a binding type with threshold $n$ and overshoot $r$ stops purchasing when $p<n$ and $nr\le(1+\tau)p$. The first inequality means that the consumer purchases the good when there is no tax, since her valuation is greater than her attention threshold $v>n$. Recall that $r=v/n$ so the second inequality states that the consumer's valuation is no greater than the amount she has to pay for the good including the tax. Define the switching indicator by
\[
e_n(r;p,\tau):=\1\{p<n,\ nr\le(1+\tau)p\}.
\]
The indicator takes a value of one if the consumer switches (stops purchasing) due to the tax and zero otherwise.

For a candidate overshoot distribution $F_R$, the switching probability and mean excess burdens are
\begin{align*}
s(p,\tau)
&=\sum_{n\in\mathcal N_I}\widehat w_{0,n}
\int e_n(r;p,\tau)\,\mathrm dF_R(r),\\
\mathrm E[\mathrm{EB}_{\mathrm{TRJ}}(p,\tau)]
&=\sum_{n\in\mathcal N_I}\widehat w_{0,n}
\int (n-p)e_n(r;p,\tau)\,\mathrm dF_R(r),\\
\mathrm E[\mathrm{EB}_{\mathrm{APT}}(p,\tau)]
&=\sum_{n\in\mathcal N_I}\widehat w_{0,n}
\int (nr-p)e_n(r;p,\tau)\,\mathrm dF_R(r).
\end{align*}
Under the TRJ interpretation, the surplus of an eliminated purchase is $n-p$ because the untaxed cutoff is interpreted as willingness to pay. Under RUM--APT, the lost surplus is $nr-p$, which equals $v-p$. Multiplying mean excess burden by 100 gives cents per binding-threshold observation, as reported in Table~\ref{tab:welfare_ratio}.

We obtain bounds by minimizing and maximizing each expression over the partially identified overshoot distributions in Section~\ref{sec:overshoot_pi}. We also minimize and maximize the ratio of mean RUM--APT to mean TRJ excess burden, evaluating both means under the same candidate distribution. 

The switching and mean excess burden bounds are computed by linear programming. The ratio bounds are computed through successive linear programs after verifying that mean TRJ excess burden is strictly positive over the feasible set.

\end{document}